\documentclass[preprintnumbers,pre,twocolumn]{revtex4}

\usepackage{amsmath}
\usepackage{graphicx,caption,subcaption,float}
\usepackage[symbol]{footmisc}
\usepackage{multirow,array}

\usepackage{siunitx}

\begin{document}

\bibliographystyle{prsty} % Choose Phys. Rev. style for bibliography

\author{Korosh Mahmoodi$^{1}$\footnote{Corresponding author: koroshm@andrew.cmu.edu}}
\author{Bruce J. West$^{2}$}
\author{Cleotilde Gonzalez$^{1}$}
\affiliation{1) Dynamic Decision Making Laboratory,
Department of Social and Decision Sciences,
Carnegie Mellon University,
5000 Forbes ave., Pittsburgh, PA15213 USA \\
2) Information Sciences Directorate,
Army Research Office,
Research Triangle Park, NC 27708, USA }

\begin{abstract}
We propose a model for demonstrating spontaneous emergence of collective intelligent behavior from selfish individual agents. Agents' behavior is modeled using our proposed selfish algorithm ($SA$) with three learning mechanisms: reinforced learning ($SAL$), trust ($SAT$) and connection ($SAC$). Each of these mechanisms provides a distinctly different way an agent can increase the individual benefit accrued through playing the prisoner's dilemma game ($PDG$) with other agents. The $SA$ provides a generalization of the self-organized temporal criticality ($SOTC$) model and shows that self-interested individuals can simultaneously produce  maximum social benefit from their decisions. The mechanisms in the $SA$ are self-tuned by the internal dynamics and without having a pre-established network structure. Our results demonstrate emergence of mutual cooperation, emergence of dynamic networks, and adaptation and resilience of social systems after perturbations. The implications and applications of the $SA$ are discussed.

\

\textit{Keywords\/{: Selfish Algorithm, Reinforcement Learning, Emergence of Collective Intelligence, Emergence of Network Reciprocity, Resilience}}

\end{abstract}

\title{Selfish Algorithm and Emergence of Collective Intelligence}
\maketitle

\section{Introduction}

\qquad In this paper we address the problem of resolving the paradox of cooperative behavior emerging in situations in which individuals are assumed to act solely in their own self interest, that is, selfishly. This particular paradox has a long history and has been widely studied in sociology and in the cognitive sciences. Game theory has been among the leading descriptors of normative behavior in this regard. In particular, the Prisoner's Dilemma Game ($PDG$) has been a leading metaphor for the study of the evolution of cooperative behavior in populations in which selfishness is substantially rewarded in the short-term \cite{nowak1993strategy,gonzalezetal2015}. Herein we integrate the $PDG$ behavior into a recently developed model of self-organized behavior \cite{west17}, but before we become immersed in the formalization of our proposed model, we examine the behavior of interest from a qualitative perspective.

It is nearly two centuries since Lloyd \cite{lloyd33} introduced the \textit{tragedy of the commons} scenario to highlight the limitations of human rationality in a social context. We subsequently update this scenario to modern situations in which individuals, with self-centered interests, must share a common resource. If each individual acts selfishly, without regard for the others sharing the resource, the unavoidable result will be the depletion of resources and subsequent tragedy. There is still much to be gained in reviewing the original scenario before adapting it to our present concerns. The 'commons' is an old-English term for a land area whose use is open to the public. Lloyd considered a pasture to be a commons, which is shared by a number of cattle herders, each of which retains as many cattle as possible. External factors such as wars, disease, poaching and so on, act to keep the number of men and beasts well below the maximum level that can be safely supported by the commons. The dynamic response to these disruptions is robust, but eventually, social stability ensues. However, instead of enhancing the well being of the situation, the stability inevitably leads to tragedy. How does this counterintuitive development arise?

In Lloyd's argument, the assumption is made that all herdsmen act rationally. On the socially stable commons, rationality leads to a herdsman acting in his own self-interest in order to maximize the gain made from grazing and selling his cattle. He reasons that, on the downside, to grow his herd incurs certain costs, the purchase price and the incremental cost of overgrazing, by adding another animal to his herd. On the upside, the overgrazing cost is shared by all the herdsmen on the commons. Consequently, since he does not share with others the profit of selling this additional animal, it makes economic sense for him to add the animal to his herd. This argument is valid for adding additional animals as well. This same conclusion is independently reached by each of the other herdsmen on the commons, since they are equally rational; resulting in tragedy, since with the growing herds the commons grazing capacity will inevitably be destroyed over time. The herdsman are trapped in a rational system that logically compels them to increase, without limit, the size of their herd and to do this in the face of limited resources. Disaster and failure is the end point of this logical sequence in which all participants act in their own self-interest.

One can find a number of weak points in this arcane argument, which we pose here as questions. Do people always (or ever) behave strictly rationally? Do individuals ever act independently of what others in their community are doing? Do people typically completely disregard the effects of their actions on the behavior of others? We subsequently address each of these questions and others in a more general context. For the moment it is sufficient to point out that the "straw man" of the tragedy of the commons is an example of linear logical thinking and its deadly flaw is the unrealistic way it is applied to resolving paradoxes in our complex world.

\paragraph{The Altruism Paradox}
Altruism is one concept that was missing from Lloyd's tragedy of the commons discussion, but which seems to be no less important than the notion of selfishness, which was foundational to his argument. Moreover, if selfishness is entailed by rationality in decision making, as he maintained, then altruism must be realized through an irrational mechanism in decision making. In point of fact, the competition between the two aspects of human decision making \cite{ariely08, kahneman2011thinking}, the rational and irrational, may well save the commons from ruin.

Let us consider the altruism paradox ($AP$), which identifies a self-contradictory condition regarding the characteristics of species. This dates back to Darwin's recognition that some individuals in a number of species act in a manner that although helpful to other members of the species, may jeopardize their own survival. Yet this property is often characteristic of that species. He also identified such altruism as contradicting his theory of evolution, the natural selection \cite{darwin71}. Darwin also proposed a resolution to this problem by speculating that natural selection is not restricted to the lowest element of the social group, the individual, but can occur at all levels of a biological hierarchy, which constitutes multilevel selection theory \cite{wilson07}.

As discussed by West et al. \cite{west19b} the theory of sociobiology was developed in the last century for the purpose of, at least in part, explaining how and why Darwin's theory of biological evolution is compatible with sociology. Wilson and Wilson \cite{wilson07} were able to demonstrate the convergence of scientific consensus on the use of multilevel selection theory to resolve the $ AP$ in sociobiology. Rand and Nowak  \cite{rand2013human} emphasize that natural selection suppresses the development of cooperation unless it is balanced by specific counter mechanisms. Five such mechanism found in the literature are: multilevel selection, spatial selection, direct reciprocity, indirect reciprocity and kin selection. In their paper they discuss models of these mechanisms, as well as, the empirical evidence supporting their existence in situations where people cooperated, but in the context of evolutionary game theory. The resolution of such fundamental problems as the $AP$ may be identified at the birth of each of the scientific disciplines; new perspectives designed to ignore large domains of science thought to be irrelevant within the narrow confines of the new discipline. The complexity of a given phenomenon can therefore be measured by the number of different disciplines interwoven to describe its behavior.

\paragraph{Prisoner's Dilemma Game}

\qquad The $PDG$ dates back to the early development of game theory \cite{RapoportChammah65}, and is a familiar mathematical formulation of the essential elements of many social situations involving cooperative behavior. $PDG$s are generally represented with a payoff matrix that provides payoffs according to the actions of two players (see Table~\ref{AAA}). When both players cooperate, each of them gains the payoff $R$, and when both players defect, each of them gains $P$. If in a social group agents $i$ and $j$ play the $PDG$, when $i$ defects and $j$ cooperates, $i$ gains the payoff $T$ and $j $ gains the payoff $S$ and when the decisions are switched, so too are the payoffs. The constraints on the payoffs in the $PDG$ are $T>R>P>S$ and $S+T<2R$. The temptation to defect is established by setting the condition $T>R$.

The dilemma arises from the fact that although it is clear that for a social group (in the short-term) and for an individual (in the long-term) the optimal mutual action is for both to cooperate, each individual is tempted to defect because that decision elicits the higher immediate reward to the individual defecting. But, assuming the other player also acts to selfishly maximize her own benefit, the pair will end up in a defector-defector situation, having the minimum payoff $P$ for both players. How do individuals realize that cooperation is mutually beneficial in the long-term? This question has been answered by many researchers, at various levels of inquiry, involving pairs of agents \cite{gonzalezetal2015, Moisanetal2018}, as well as, larger social networks \cite{nowak1993strategy}. Research suggests that, at the pair level, people dynamically adjust their actions according to their observations of each others' actions and outcomes; at the complex dynamic network or societal level, this same research suggests that the emergence of cooperation may be explained by \emph{network reciprocity}, whereby individuals play primarily with those agents with whom they are already connected in a network structure. The demonstration of how social networks and structured populations with explicit connections foster cooperation was introduced by Nowak and May \cite{nowak1992evolutionary}. Alternative models based on network reciprocity assume agents in a network play the $PDG$ only with those agents to whom they have specific links. Agents act by copying the strategy of the richest neighbor, basing their decisions on the observation of the others' payoffs. Thus, network reciprocity depends on the existence of a network structure (an already predefined set of links among agents) and on the awareness of the behavior and payoffs of interconnected agents.

\begin{table}
    \setlength{\extrarowheight}{2pt}

\caption{The general payoffs of $PDG$. The first value of each pair is the payoff of agent $i$ and the second value is the payoff of the agent $j$.}  
    \begin{tabular}{cc|c|c|}
      & \multicolumn{1}{c}{} & \multicolumn{2}{c}{Player $j$}\\
      & \multicolumn{1}{c}{} & \multicolumn{1}{c}{$C$}  & \multicolumn{1}{c}{$D$} \\\cline{3-4}
      \multirow{2}*{Player $i$}  & $C$ & $(R,R)$ & $(S,T)$ \\\cline{3-4}
      & $D$ & $(T,S)$ & $(P,P)$ \\\cline{3-4}
    \end{tabular}

\label{AAA}

  \end{table}

\paragraph{Empirical evidence of emergence of cooperation}
Past research has supported the conclusion that the survival of cooperators requires the observation of the actions and/or outcomes of others and the existence of predefined connections (links) among agents. Indeed, empirical work suggests that the survival and growth of cooperation within the social group depends on the level of information available to each agent \cite{MartinGonJuvLeb2014}. The less information about other agents available, the more difficult it is for cooperative behavior to emerge \cite{RapoportChammah65, MartinGonJuvLeb2014}. On the other hand, other experiments suggest that humans do not consider the payoffs to others when making their decisions, and that a network structure does not influence the final cooperative outcome \cite{fischbacher2001people}. In fact, in many aspects of life, we influence others through our choices and the choices of others affect us, but we are not necessarily aware of the exact actions and rewards received by others that have affected us. For example, when a member of society avoids air travel in order to reduce their carbon footprint, s/he might not be able to observe whether others are reducing their air travel as well, yet they rely on decisions others make, influencing the community as a whole. Thus, it is difficult to explain how behaviors can be self-perpetuating even when the source of influence is unknown \cite{MartinGonJuvLeb2014}.

These empirical observations support an important hypothesis emerging from the work presented herein. Our hypothesis is that mutual cooperation emerges and survives, even when the social group consists exclusively of selfish agents and there is no conscious awareness of the payoffs to other agents. Moreover a network structure can emerge dynamically from the connections formed and guided by individual selfishness. Note that this is the dynamics \underline{of} a network, as distinct from the more familiar dynamics \underline{on} a network. The dynamics on a network assumes a static network structure, as in evolutionary game theory, whereupon the strengths of the links between agents may change, but the agents sit at the nodes of the network and interact with the same nearest neighbors. On the other hand, the dynamics of a networks makes no such assumption and the dynamics consist of the formation and dissolution of links between any two agents within the social group. 

\section{Contributions}
Herein, we aim to clarify how collective intelligence emerges without explicit knowledge of the actions and outcomes of others, and in the absence of a predefined network structure linking agents within a society. We introduce an algorithm (the \emph{Selfish Algorithm}, $SA$) to demonstrate that collective intelligence can emerge and survive between agents in a social group, out of selfishness. We construct a computational model of the $SA$ and generate simulations of the evolution of mutual cooperation and networks out of selfishness. 

First, the $SA$ model provides a resolution of the altruism paradox ($ AP$) and shows how agents create a self-organized critical group out of individual selfish behavior, simultaneously maximizing the benefit of individual and of the whole group. Second, the $SA$ model demonstrates how \textit{adaptation} can naturally emerge from the same mechanisms in the model. Adaptation is an important property of living things, which allows them to respond to a changing environment in a way that optimizes their performance and survivability.

We studied four systems governed with different learning mechanisms proposed by the $SA$, where the behavior of the agents is modeled using three modes of learning: reinforced learning ($SAL$), trust ($SAT$), and connection ($SAC$). The first system studied has only the $SAL$ mechanism active. The other three systems are a combination of $SAL$ and $SAT$ ($SALT$), $SAL$ and $SAC$ ($SALC$), and the combination of all the mechanisms ($SALTC$). Next, we tested the sensitivity of the collective behavior that emerged from these systems to changes in the social makeup by modifying a fraction of the agents having them exchanged with zealots at a given time. A zealot is an agent that will not change its decision regardless of the payoff. The comparison of the mutual cooperation of the systems versus the number of the zealots is a measure of resilience, or robustness (i.e., the resistance of the collective behavior to perturbations).

The computational results show that these systems can be ranked from strongest to weakest in their resilience as $ $SALTC$ > $SALC$ > $SALT$ > $SAL$ $. The $SALC$ and $SALTC$ systems have a high resilience because using $SAC$ in the decision making process enables agents to learn to avoid pairing with the zealots. This way of pairing is different from the popular mechanism of preferential attachment \cite{barabasi1999emergence}, which uses rules according to which an agent would have a higher chance of linking with other agents that already have many links, those being agents with a substantial reputation. In contrast, the $SA$ demonstrates that such chances to connect to other agents emerge dynamically, according to the experienced benefits that the other agent brings to the individual's own benefit ($SAC$). This high adaptability of $SA$ shows that each agent can be observed to act as a local leader, making decisions based on changes in its environment toward maximizing its own benefit and, at the same time, the local decisions made by agents also affect the choices taken by the system as a whole, leading it to its optimal state.

\paragraph{Network reciprocity and survival of cooperators}

\begin{figure*}[bt]
\centering
\includegraphics[width=15cm]{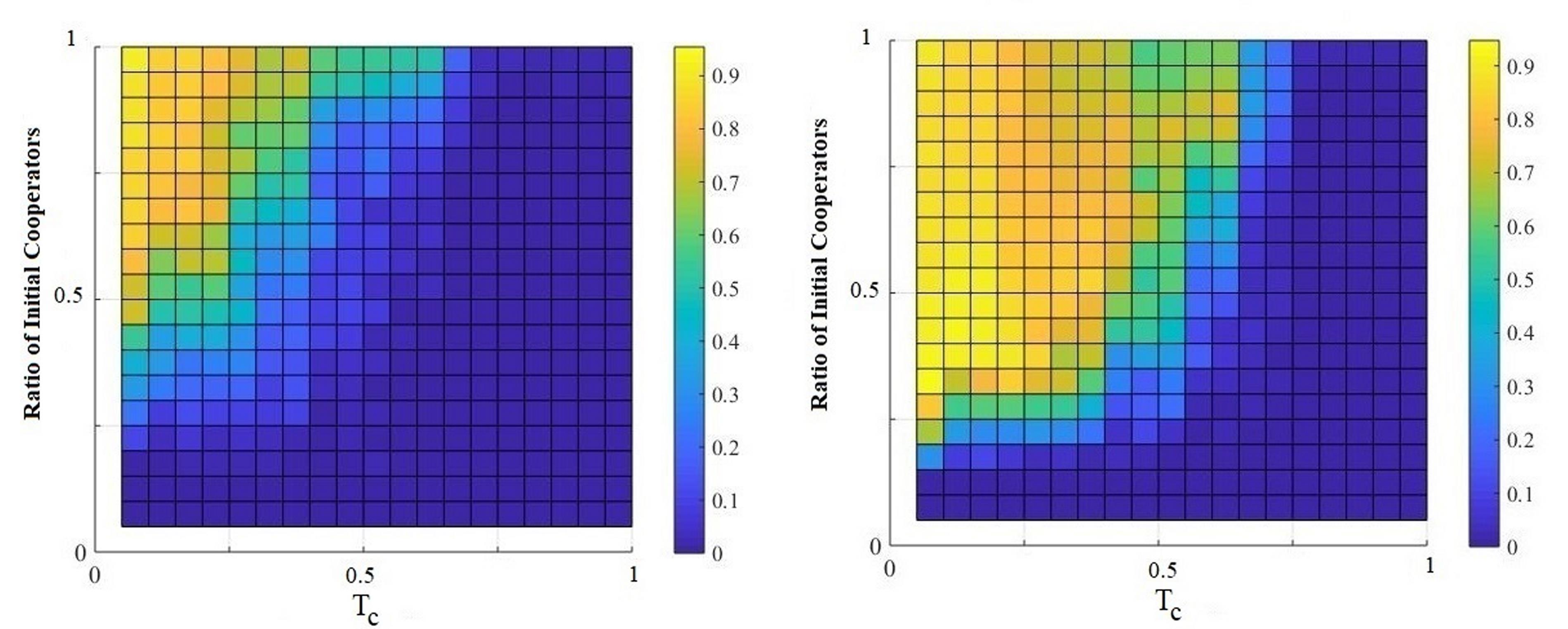}
\caption{The colors on the left and right panel show the average of the ratio of Mutual Cooperation ($RMC$) at time 100, when each agent played the $PDG$ for 100 times with its eight nearest neighbors, and updated its decision by imitating the decision of the richest neighbor (including themselves). The agents were located on a regular two-dimensional lattice of size 10$\times $10 (left panel) and  30$\times $30 (right panel). The horizontal axis of the panels shows the degree of temptation to cheat experienced by the agents  and the vertical axis is the ratio to initial cooperators on the lattice. 100 ensembles were was used to evaluate the average $RMC$ at time 100.}
\label{NowakPhase}
\end{figure*}

\begin{figure*}[bt]
\centering
\includegraphics[width=15cm]{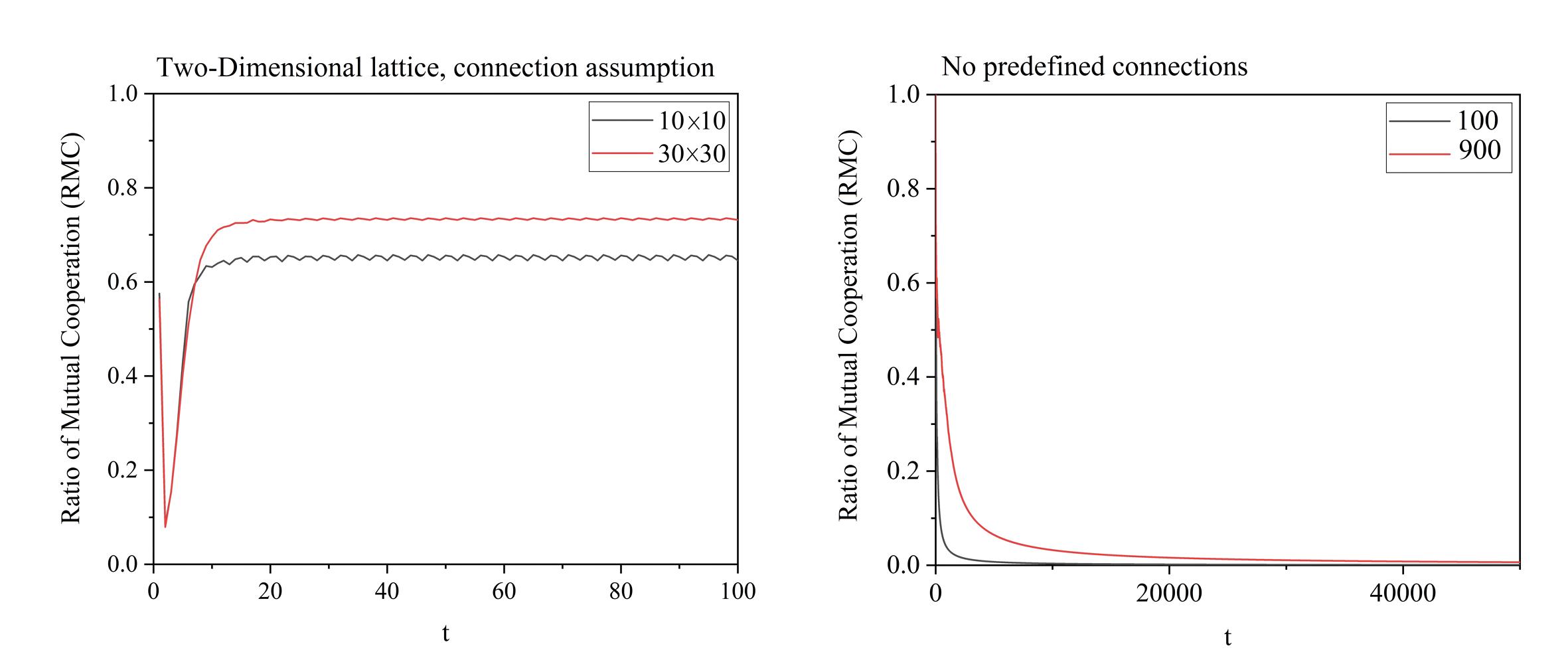}
\caption{Left panel: The time evolution of the Ratio of Mutual Cooperation ($RMC$) for agents located on a regular two-dimensional lattice with 10$\times $10  (black curve) and 30$\times $30 agents (red curve). Initially 75\% of agents randomly picked to be cooperators and at each time agents played $PDG$ with $T_{c} = 0.25$ with their 8 nearest neighbors and imitated the decision of its richest neighbor (including themselves). Right panel: The time evolution of the $RMC$ for the agents with similar condition as those adopted in the left panel except that at each time two agents are picked randomly and played $PDG$ (no predefined connections). The curves are averaged over 100 realizations.}
\label{NoNet}
\end{figure*}

\qquad To put our contributions in perspective, let us highlight the main differences between the $SA$ model and those of previous studies. First, the works following the strategy of the Nowak-May agents \cite{nowak1992evolutionary} typically compare their individual payoff with the payoffs of the other agents when making their next decision. One way to realize this strategy is to adopt the decision of the most successful neighbor. In contrast, all the updates in the $SA$ are based on the last two payoffs to the self-same individual. In other words, $SA$ agents do not rely on information about the other members of society to determine its evolution. $SA$ agents are connected to other agents, only indirectly, through their local payoffs in time. The most recent payoff to the individual is the result of its interaction with its most recent partner, but its previous payoff, used for tuning its tendencies (presented below in Equation \ref{prob}), might be the result of playing with a different partner. Second, the Nowak-May agents \cite{nowak1992evolutionary} are assumed to be located at the nodes of a lattice and to be connected by links, as they were also  assumed to be in the decision making model ($DMM$)  \cite{west19b}. For example, because of social interactions these lattice networks are regularly used to model complex networks and have historically been assumed to mimic the behavior of real social systems. These studies emphasized the important role of network complexity in sustaining, or promoting, cooperation \cite{tomassini2007social}. This was accomplished without  explicitly introducing a self-organizing mechanism for the network formation itself.

To show the importance of the specific connection of agents on a lattice in Nowak and May \cite{nowak1992evolutionary} model, we replicated their simulation work. In this simulation the agents are located on a regular two-dimensional lattice, each having eight nearest neighbors to play the $PDG$. There is an agent located at each node of the lattice and they each update their decision at each round of the computation by imitating the decision of the richest agent in their neighborhood (including themselves). Figure~\ref{NowakPhase} shows our replication of their work. The colors on the panels indicate the average of the Ratio of Mutual Cooperation ($RMC$) which sustained between the agents located on the 10$\times$10 lattice (left panel) and on the 30$\times$30 lattice (right panel) after each agent played 100 times with its eight nearest neighbors. The yellow areas correspond to high $RMC$ which happened for low temptation to cheat $T_{c}$ (i.e., agent's selection of the Defect action in the $PDG$) and  high initial number of cooperators. The yellow area is more extended in the case of agents located on the larger lattice of size 30$\times$30 (right panel).

To explicitly show the importance of Nowak and May's assumption of the lattice structure for survival of mutual cooperation, we ran their model and compared time evolution of the RMC between these agents in the case where the agents were paired up randomly. The left panel of Figure~\ref{NoNet} shows the ensemble average of 100 simulations for 100 agents (black curve) and 900 agents (red curve) connected on a two-dimensional lattice. At each time round, each agent plays the $PDG$, with a low temptation to cheat $T_{c}(=0.25)$, with all its eight neighbors. Initially 75\% of the agents were cooperators, but the $RMC$ evolved and sustained in both cases. The right panel of Figure~\ref{NoNet} shows the results of the same experiment, but selecting the interacting pairs randomly (no lattice structure is assumed for the agents). As shown in the figure, the $RMC$ vanishes in the absence of the network structure.

As we demonstrate next, the $SA$ does not rely on a lattice network, yet, mutual cooperation is subsequently shown to emerge and survive the induce of perturbations. In fact, we show that the network reciprocity can be interpreted to be a byproduct of $SA$, emerging out of the selfishness of the $SA$ agents.

\section{The Selfish Algorithm ($SA$)}

\qquad The selfish algorithm ($SA$) proposed in this research belongs to a family of agent-based models in modern evolutionary game theory  \cite{adami2016evolutionary}. Specifically, $SA$ represents a generalization of a model introduced by Mahmoodi et al. \cite{west17}  which is a coupling of $DMM$ with evolutionary game theory put into a social context. This earlier introduction led to the self-organized temporal criticality ($SOTC$) concept and the notion that a social group could and would adjust their behavior to achieve consensus, interpreted as a form of group intelligence, as had been suggested by the collective behavior observed in animal studies \cite{couzin2007collective}. The $SOTC$  agents,   like Nowak and May's agents \cite{nowak1992evolutionary}, assumes a pre-existing network between the agents which use the $DMM$ (belonging to the Ising universality class of models for decision making \cite{li2018ising}). $SA$ overcomes this limitation, resulting in a general model for emergence of collective intelligence and a complex dynamic network.

The general notion that the $SA$ adopted from Mahmoodi et al. \cite{west17} is that an agent $i$ makes a decision according to the change in a cumulative tendency (i.e., preference). The change in this cumulative tendency, $\Delta$, is a function of the last two payoffs of agent $i$ who played with agent $j$ and $k$:
\begin{equation}
\Delta _{i,jk}=\chi \frac{\Pi _{ij}(t)-\Pi _{ik}(t-1)}{|\Pi _{ij}(t)|+|\Pi
_{ik}(t-1)|},  \label{prob}
\end{equation}
where $\chi $ is a positive number that represents the sensitivity of the agent to its last two payoffs. The quantity $\Pi_{ij}(t)$ is the payoff to the agent $i$ played with agent $j$ at time $t$ and $\Pi _{ik}(t-1)$ is the payoff to the agent $i$ played with agent $k$ at time $t-1$. We used the $S$ value in the $PDG$ that is $> 0$ and assumed $\Delta _{i,jk}= 0$ when the denominator of Eq. (~\ref{prob}) is zero.

A simplified version of the $SA$ was previously introduced elsewhere \cite{mahmoodi19}, and an abbreviated version of the algorithm and its mathematical details are included in Appendix A. In the first cycle of the $SA$, a pair of randomly selected agents $i$ and $j$ "agree" to play. Only one pair of agents play at each time cycle. Each of the agents of a pair engages in three decisions; each decision is represented in a learning mechanism (a cumulative propensity lever, ranging from 0 to 1) by which the agent can gain information about the social environment: learning from their own past outcomes (Selfish Algorithm Learning, $SAL$), trust the decision of other agents (Selfish Algorithm Trust, $SAT$), and make social connections that are beneficial (Selfish algorithm-based connection, $SAC$). Each of these levers is updated according to the agent's self interest (selfishness): a decision is made according to the state of each lever, and the lever is updated according to $\Delta$ as formalized in Eq. (~\ref{prob}).

In the $SAL$ mechanism, an agent decides to cooperate ($C$) or defect ($D$) while playing the $PDG$ with the paired partner, according to the state of the cumulative propensity of playing $C$ or $D$. The cumulative propensity of the agent to pick $C$ or $D$ increases (or decreases) if it's payoff is increased (or decreased) with respect to its previous payoff, as per Eq. (~\ref{prob}). The updated cumulative tendency to play $C$ or $D$ with the same agent is used for the next time agent $i$ is paired with the same agent $j$.  Our simulation results show that the $SAL$ mechanism attracts the agents toward mutual cooperation.

In the $SAT$ decision, an agent decides to rely on the decision of its partner, instead of using its own decision made using $SAL$, according to the state of the cumulative propensity of "trusting" the other's decision or not. Each $SAT$ agent can tune this propensity to rely on its partner's decision according to $\Delta$ as formalized in Eq. (~\ref{prob}). The $SAT$ agent increases (or decreases) its tendency to trust its partner's decision if it increased (or decreased) its payoff with respect to its previous payoffs. Our simulation results show that the $SALT$ (both $SAL$ and $SAT$ active) mechanism amplifies the mutual cooperation between the agents regardless of the value of the incentive to cheat $T_{c}$.

Finally, the $SAC$ is a decision of an agent to pick the partner with whom to play in each round. Each $SAC$ agent can tune its propensity of selecting its partner according to $\Delta$ as formalized in Eq. (~\ref{prob}). A $SAC$ agent increases (or decreases) its tendency of playing with the same partner if the payoff received after playing with that partner is higher (or lower) than the agent's own previous payoffs. Our simulation results show that a network of connections emerge over time between $SAC$ agents, and that this network amplifies the mutual cooperation between agents.

\section{Simulation Results}
We studied  $20$ agent system and set the parameters for the calculations as follows: Initially all the agents were defectors, had  payoff of zero, a propensity of 0.9 to remain a defector,  propensity of 0 to trust the partner's decision, and equal propensity of 1/(20-1) to connect with each of the other agents. We set the sensitivity coefficient to $\chi =200$. The parameter $\chi$  controls the magnitude of the changes of the cumulative tendencies at each update where we set the maximum and minimum of the cumulative tendencies to be 1000 and 0, respectively. As a boundary condition, if the updated  cumulative tendency goes beyond 1000 (or below 0) then it is set back to 1000 (or 0). A smaller $\chi $, with respect to the maximum value of the cumulative tendency, slows down the process of learning, but it does not qualitatively change behavior, but merely scales the dynamical properties of the system. The time for simulations are picked to be sufficiently long to show the asymptotic behavior of the systems.
 
The payoff matrix used in the simulations is shown in Table~\ref{Pay} as suggested by Gintis \cite{gintis2009bounds}: $R=1$, $P=0$ and $S=0$. So, the maximum possible value of $T=1+T_{c}$ is 2. We selected the value $T_{c}=0.9$ (unless otherwise explicitly mentioned) which provides a strong incentive to defect. This simple payoff matrix has the spirit of the $PDG$ and  allows us to emphasize the main properties of the SA model.

\begin{table}
    \setlength{\extrarowheight}{2pt}

\caption{The payoffs of the $PDG$. The first value of each pair is the payoff of agent $i$ and the second value is the payoff of the agent $j$.}  
    \begin{tabular}{cc|c|c|}
      & \multicolumn{1}{c}{} & \multicolumn{2}{c}{Player $j$}\\
      & \multicolumn{1}{c}{} & \multicolumn{1}{c}{$C$}  & \multicolumn{1}{c}{$D$} \\\cline{3-4}
      \multirow{2}*{Player $i$}  & $C$ & $(1,1)$ & $(0,1+T_{c})$ \\\cline{3-4}
      & $D$ & $(1+T_{c},0)$ & $(0,0)$ \\\cline{3-4}
    \end{tabular}

\label{Pay}

  \end{table}

\subsection{Emergence of mutual cooperation by $SAL$}

\begin{figure}[bt]
\centering
\includegraphics[width=8cm]{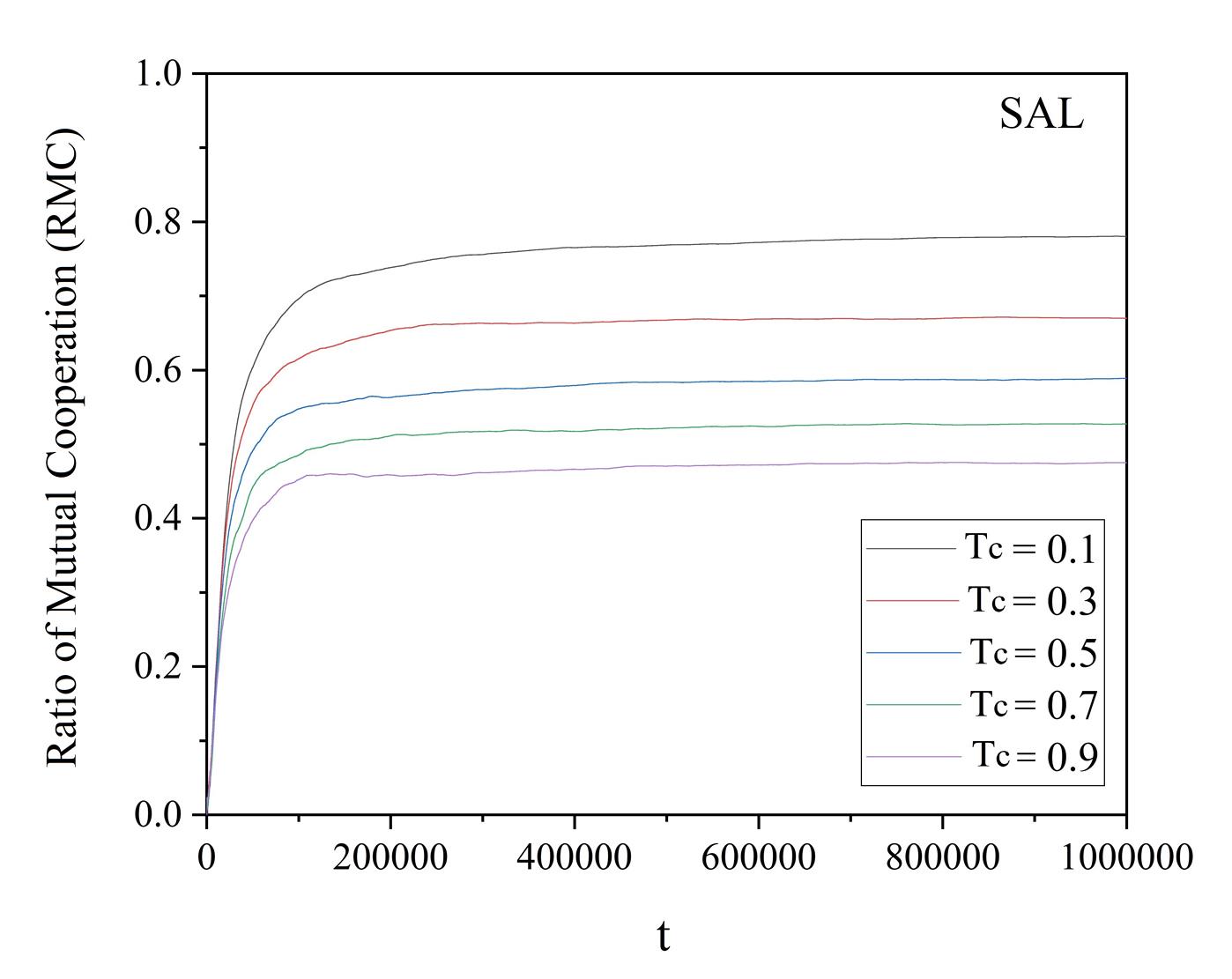}
\caption{Each curve shows the time
evolution of the Ratio of Mutual Cooperation ($RMC$) for $20$ agents used $SAL$ and played the $PDG$ with $T_{c}=0.1,0.3,0.5,0.7,0.9$ corresponding to the black, red, blue, green and purple curves, respectively.}  
\label{CCSAL}
\end{figure}

\begin{figure*}[th]
\centering
\includegraphics[width=15cm]{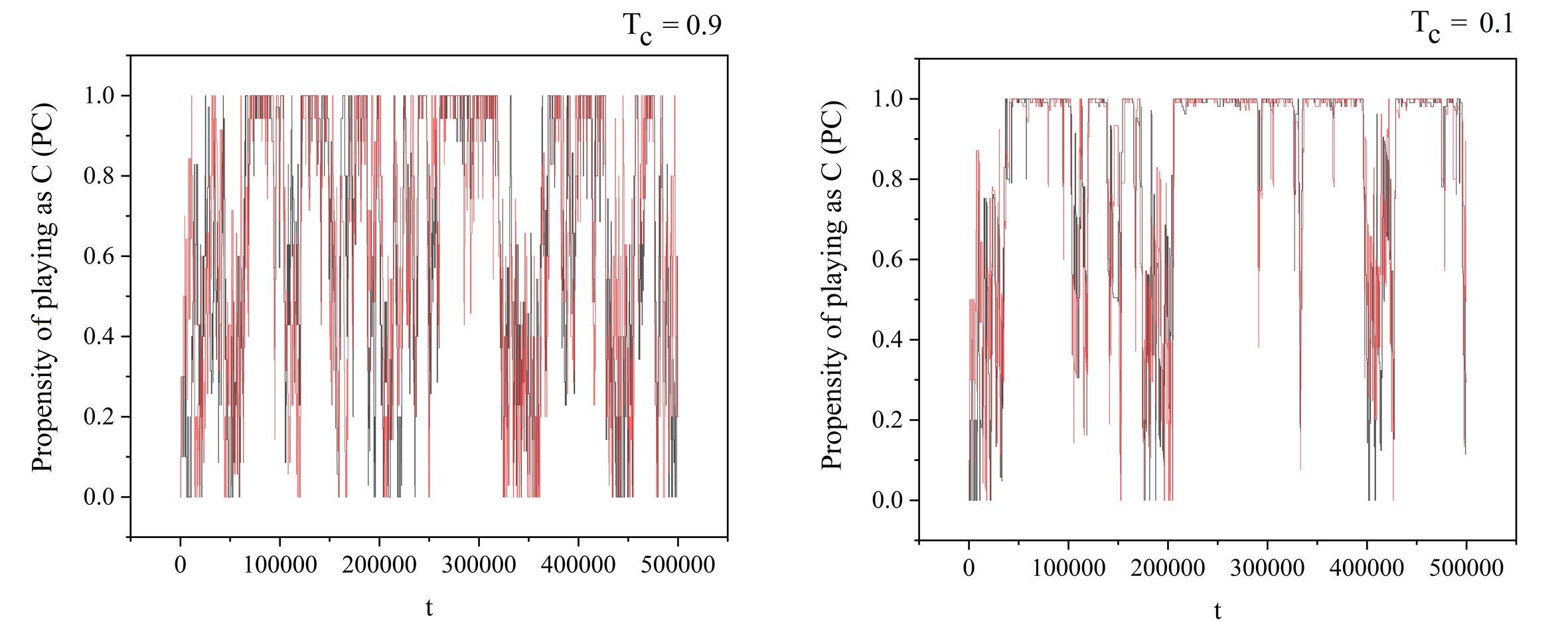}
\caption{The curves show the time evolution of the propensity of agent 1 to play as $C$ with agent 2 (red curve) ($PC_{12}$) and the propensity of agent 2 to play as $C$ with agent 1 (black curve) ($PC_{21}$). Agents 1 and two are among $20$ agents using $SAL$ to update their decisions and had $T_{c}=0.9$ (left panel) and $T_{c}=0.1$ (right panels).}
\label{PropensitySAL}
\end{figure*}

In this section we demonstrate that the learning mechanism of $SAL$ attracts agents, playing the $PDG$, to a state of mutual cooperation. Figure~\ref{CCSAL} shows the time evolution of the $RMC$ in simulations where the 20 agents, randomly partnered in each cycle, used only the $SAL$ mechanism to update their decisions. The $RMC$ increased  from zero and asymptotically saturates to a value that depends on the value of the temptation to cheat $T_{c}$. The smaller the $T_{c}$ value the higher the saturated $RMC$ value and the greater the size of the collective cooperation. Using $SAL$, each agent learns, by social interactions with the other agents, to modify its tendency to play $C$ or $D$. These agents are connected through their payoffs. Each agent compares its recent payoff with its previous one, which, with high propensity, earned by playing with a different agent. This mutual interaction between the selfish agents led them to form an intelligent group that learns the advantage of mutual cooperation in a freely connected environment whose dynamics are represented by the $PDG$.

To explain how the collective cooperation emerges, we looked into the evolution of the propensities to cooperate between two selected agents (among 20). Figure~\ref{PropensitySAL} depicts the time evolution of the propensity of two agents to pick cooperation ($C$) in the $PDG$. The propensity of agent 1 to cooperate with agent 2 (drawn in red), and the propensity of agent 2 to cooperate with agent 1 (drawn in black).  Generally, we observe a high correlation between the propensities to cooperate between the two agents.

The left panel presents these propensities when the temptation to defect is $T_{c}=0.9$, and the right panel presents the propensities when the temptation to defect is $T_{c}=0.1$. The correlation between the propensity of the two agents  is 0.65 for $T_{c}=0.9$ and 0.84 for $T_{c}=0.1$. The correlation between the agents is the result of the learning process between them. Let's assume $T_{c} = 0.5$ and that agent 1 played $C$ and agent 2 played $D$ at time $t$ which results in payoffs of 0 and 1.5 for them, respectively. Comparing its payoff with its previous payoff, earned playing with other agent (= 1.5, 0, 1 or 0), agent 1 would change its accumulative tendency to play $C$ for next time it is randomly paired with player 2 by $-\chi$, 0, $-\chi$ or 0. This means that agent 1 reacts to the defective behavior of agent 2 by tending to behave $D$ and consequently agent 2 wouldn't continue to have the advantage of a cooperative environment. On the other hand, agent 2 would change its accumulative tendency  to play $D$ with agent 1 by 0, $\chi$, $0.2\chi$ or $\chi$. This means agent 2 would like to play as $D$ next time it pairs with agent 1. So, both agents learn to play $D$ with one another, leading to coordination state of $DD$ where both get payoff of 0. Such pairs compare this payoff (= 0) with their previous payoff, played with other agent, and would change their accumulative tendency to play $D$ by $-\chi$, 0, $-\chi$ or 0 which shift their future decisions toward the coordination state of $CC$. These agents would change their tendency to play $C$ by $-0.2\chi$, $\chi$, 0 or $\chi$ which favors their stay as $C$ toward one another. However, because of the time to time change of  $-0.2\chi$ (depending on the value of $T_{c}$) in the accumulative tendency to play $C$ with other pairs, there is always a chance for agents to play $D$ for a while, before they are pulled back by other agents to behave as $C$. This creates a dynamic equilibrium between $CC$ and $DD$ states and defines the level of emerged mutual cooperation observed in Figure~\ref{CCSAL}. Thus, the dual dynamic interaction, based on self-interest, between each agent and its environment causes the emergence of mutual cooperation between the agents who play $PDG$ and use $SAL$ to update their decisions. Note that in human experiments, humans learning from only their own outcomes without awareness of the partners' outcomes did not lead to mutual cooperation \cite{MartinGonJuvLeb2014} as the rational agents of $SAL$ can do. High correlation between the propensities of the pairs of the agents using $SAL$ means high coordination between their decisions which also can occur if the "Trust"  mechanism is active between the agents. Trust lowers the intermediate $CD$ pairings and leads the agents to coordinate and converge in the same decision. In the next section we show that combining $SAL$  with $SAT$ amplifies the $RMC$ between the agents. Our current experimental work \cite{Gonzalezetalinprep} also confirms that adding the trust mechanism in decision making experiments with of human pairs helped them to realize the advantage of mutual cooperation.

\subsection{Enhancement of Mutual Cooperation by $SALT$; Trust as a Dynamic Decision}
\begin{figure}[bt]
\centering
\includegraphics[width=8cm]{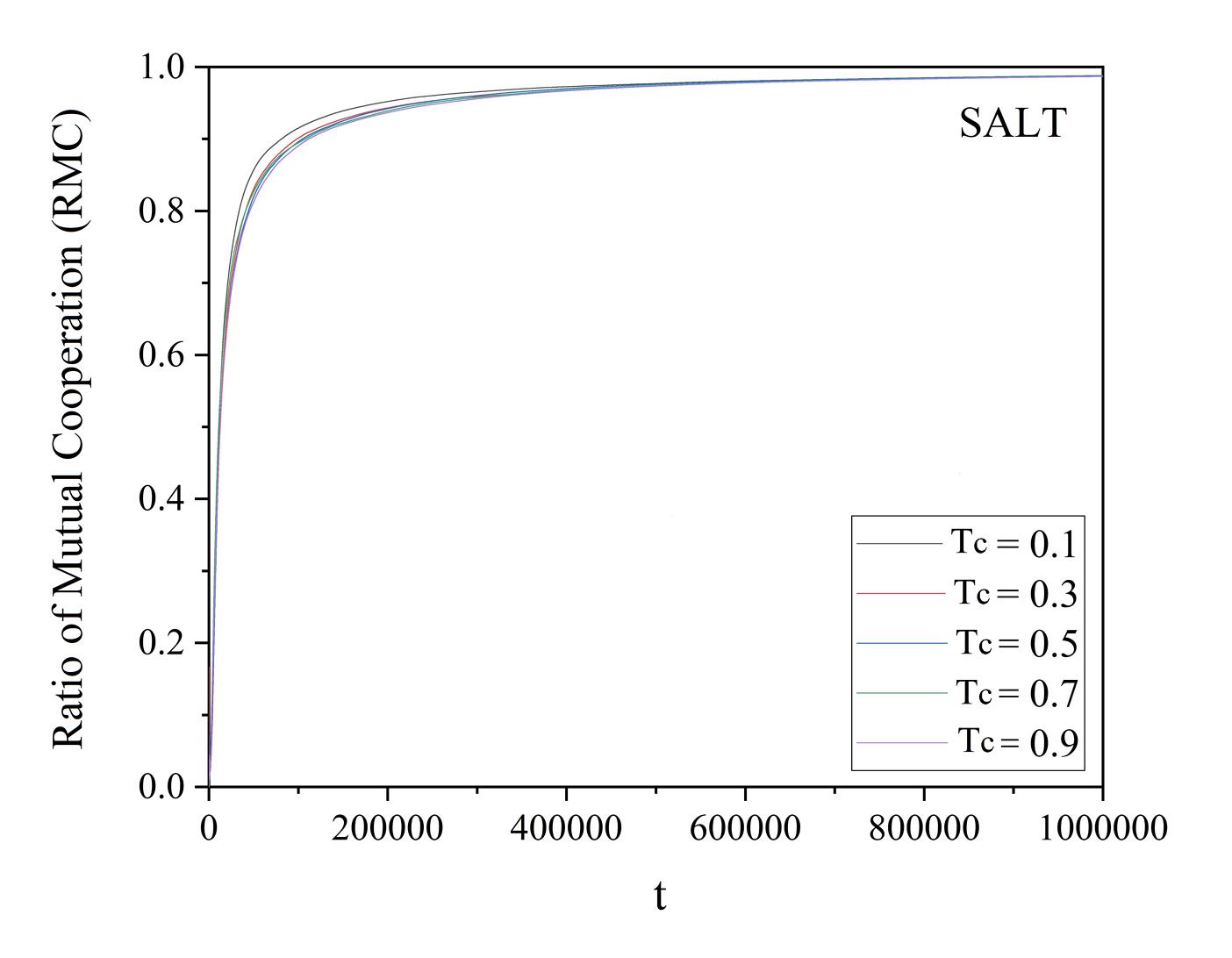}
\caption{Each curve shows the time
evolution of the Ratio of Mutual Cooperation ($RMC$) for $20$ agents using $SALT$ and playing the $PDG$ with $T_{c}=0.1,0.3,0.5,0.7,0.9$ corresponding to the black, red, blue, green, and purple curves, respectively.}  
\label{CCSALT}
\end{figure}

\begin{figure}[bt]
\centering
\includegraphics[width=8cm]{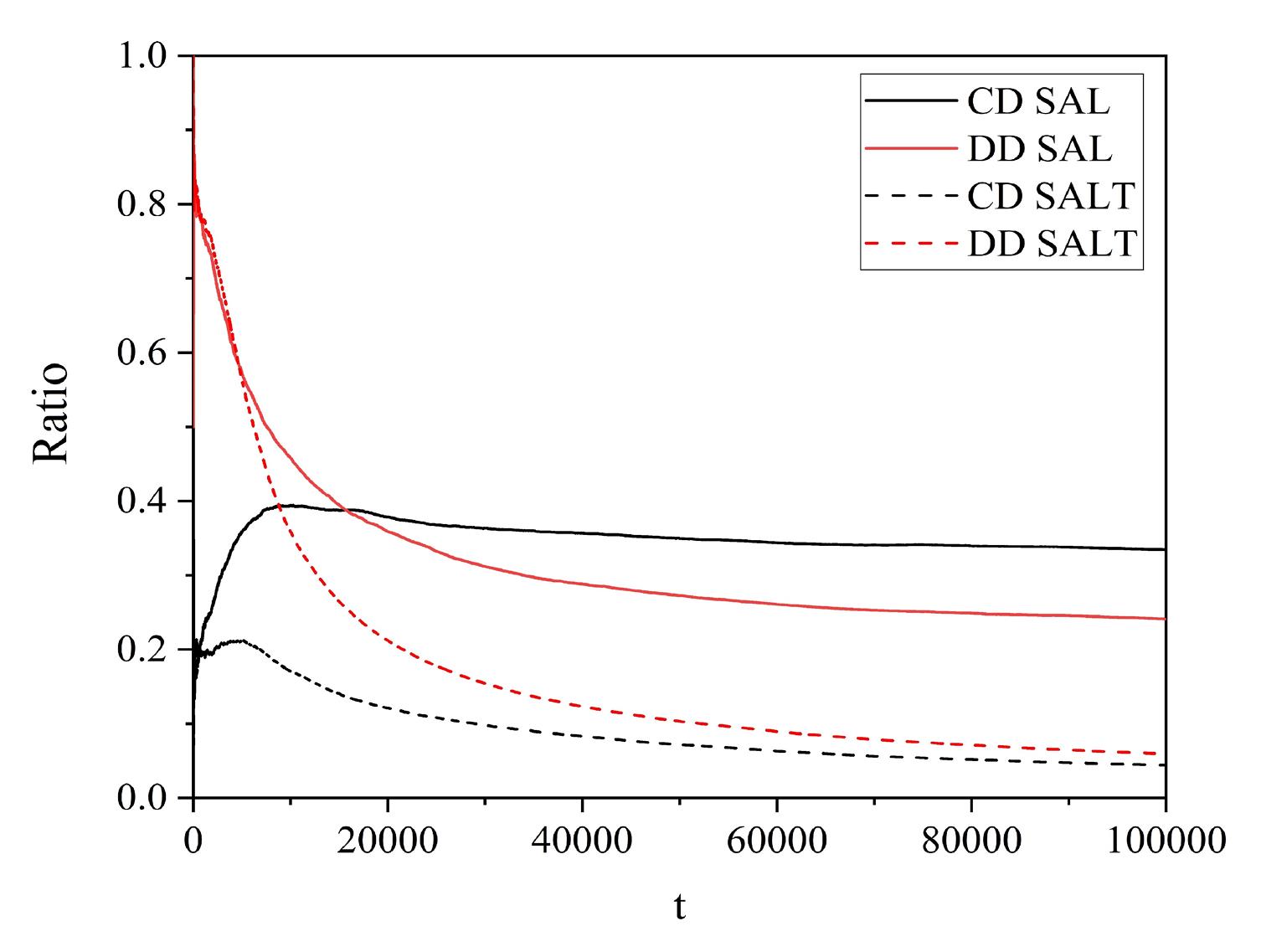}
\caption{The solid black and the solid red curves show, respectively, the time evolution of the ratio of $CD$ and $CD$ pairs played $PDG$ and used $SALT$ to make decision. The dashed black and dashed red curves show, respectively, the time evolution of the ratio of $CD$ and $CD$ pairs played $PDG$ and used $SALT$ to make decision. $M = 20$, $T_{c}=0.9$.}
\label{EffectofT}
\end{figure}

Figure~\ref{CCSALT} shows the enhancing effect on the emergence of collective cooperation when each agent is allowed to make a decision whether to "Trust" or rely on the decision made by the paired agent or not ($SAT$).

The $SALT$ mechanism also decreases the time for agents to realize the benefit of mutual cooperation. In Figure~\ref{EffectofT} we compare the ratio of $CD$ partners in a group of 20 agents playing the $PDG$ with a temptation to cheat of $T_{c}=0.9$ when using the $SALT$ model (dashed lines) and when using the $SAL$ model (solid lines). It is apparent that because of the trust mechanism, $SALT$ agents coordinate more often than they do without it and consequently avoid the formation of $CD$ pairs. We also plot the ratio of $DD$ pairs for the two systems. These curves show that $SALT$ agents learn to select $CC$ pairs over $CD$ pairs more readily than do $SAL$ agents, thereby pushing the $RMC$ up to approximately 0.9. The average correlation between the  propensity of two agents to play $C$ with one another is about 0.92 when  $T_{c}=0.9$ which is a sign of high coordination. We highlighted the difference between trust used in the literature with our dynamic trust model in Appendix C.

\subsection{Emergence of network reciprocity from $SALC$ and $SALTC$}
By activating the ability of the agents to make decisions about the social connections that are beneficial to themselves ($SAC$), we expect that an even larger and faster increase in collective cooperation. The Connection mechanism allows each agent to select a partner that helped the agent to increase its payoff with respect to its previous payoff.

\begin{figure*}[bt]
\centering
\includegraphics[width=16cm]{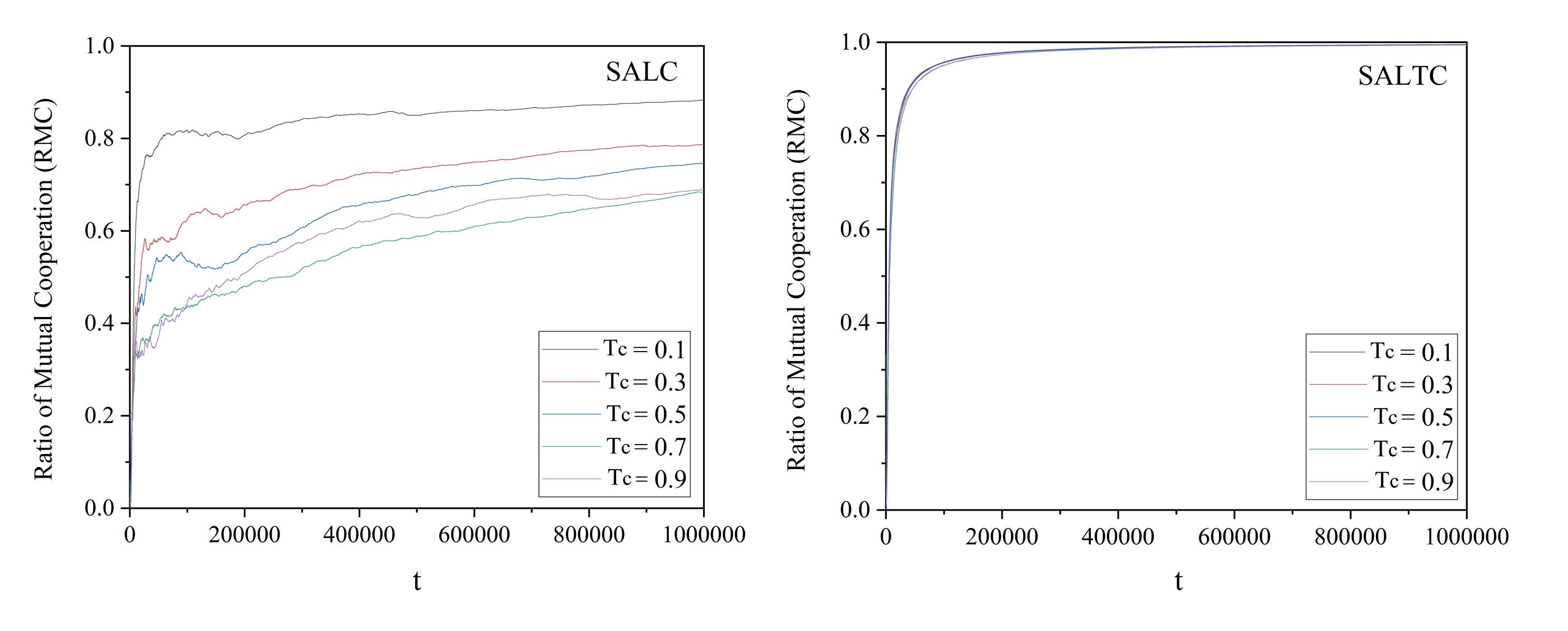}
\caption{Each curve shows the time
evolution of the Ratio of Mutual Cooperation ($RMC$) for $20$ agents used $SALC$ (left panel) or $SALC$ (right panel) and  played the $ PDG$ with $T_{c} = 0.1,0.3,0.5,0.7,0.9$ corresponding to the black, red, blue, green and purple curves, respectively.}  
\label{CCSALC}
\end{figure*}

We demonstrate the increase in the $RMC$ for a model without the Trust mechanism $SALC$ and a model with the Trust mechanism $SALTC$. The Left panel of Figure~\ref{CCSALC} shows an increase in the $RMC$ of the agents using $SALC$ (left panel), and using $SALTC$ (right panel). When comparing the $SALC$ model behavior to that of the agents without the connection mechanism (paired randomly) in Figure~\ref{CCSAL}, we observe that the dependence on the temptation to cheat $T_{c}$ is weaker. For example at time $t = 10^6$ the average $RMC$ for the agents using $SALC$ are about 0.84, 0.80, 0.75, 0.65 and 0.62 for  $T_{c} = 0.1,0.3,0.5,0.7$ and $0.9$, respectively, whereas for the agents using $SAL$  the average $RMC$ for the same conditions are about 0.76, 0.64, 0.58, 0.52 and 0.45, respectively. On the right panel of Figure~\ref{CCSALC}, we observe that using the Learning, Trust, and Connection mechanism in conjunction ($SALTC$), the level of collective cooperation is the highest with the least dependence on the temptation to cheat.

\begin{figure*}[bt]
\centering
\includegraphics[width=15cm]{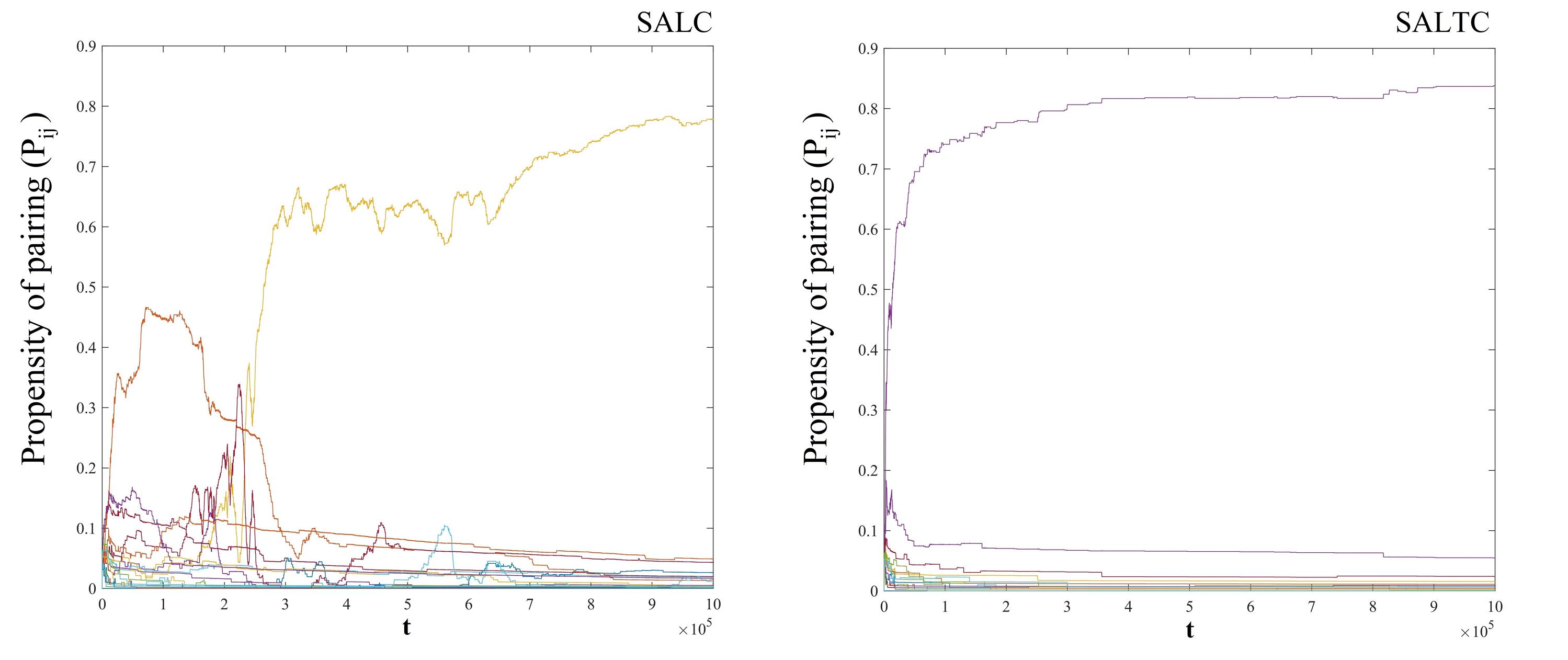}
\caption{Left and right panels show the time evolution of the propensity of pairings between agent 1 and the other 19 agents when agents used $SAC$ or $SALTC$, respectively. Agents had $T_{c}=0.9$.}
\label{Recip}
\end{figure*}

To show that the reciprocity emerged between the agents using the $SAC$ mechanism, we plotted in  Figure~\ref{Recip} the propensities of making connections between a typical agent (agent 1) and the other 19 agents where agents used $SALC$ (left panel) or $SALTC$ (right panel) to update their decisions. In both cases, these figures show that agent 1 developed a preferential partner and learned to play most of the time with one of the agents among others.

\begin{figure*}[th]
\centering
\includegraphics[width=18cm]{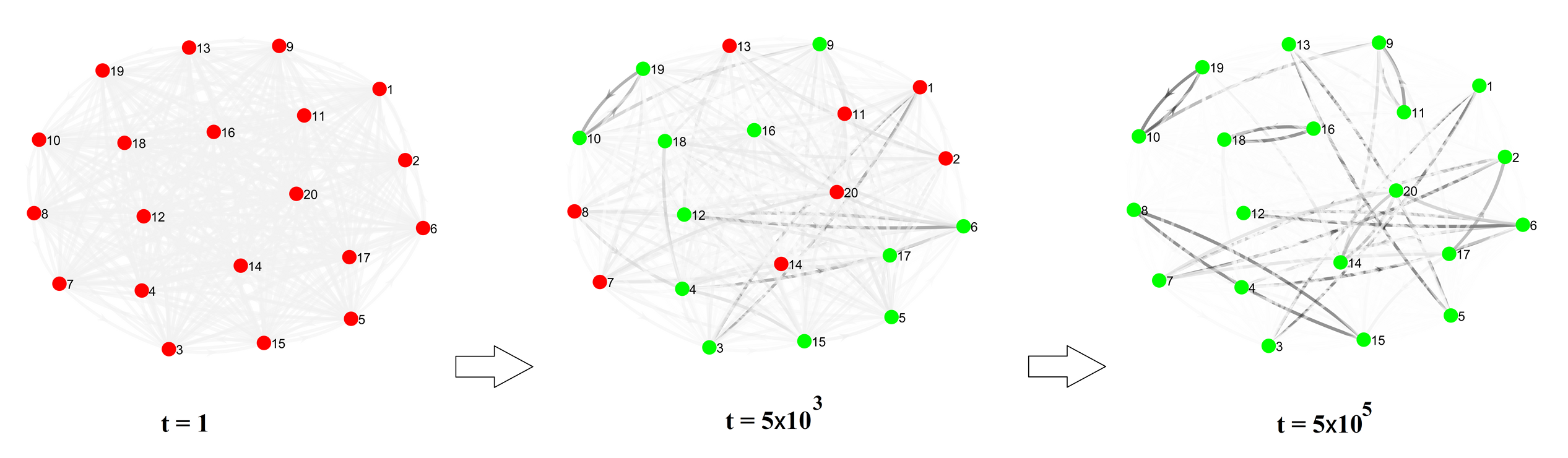}
\caption{Emergence of reciprocity from dynamics \underline{of} the network. From left to right the panels show the snapshots of the propensities of connections between 20 agents played the $ PDG$ and updating their decisions based on $SALTC$ at $t= 1$, $5\times10^{3}$ and $5\times10^{5}$, respectively. Intensity of the lines between pairs represent the magnitude of the propensity of one to another and the directions show the intensity belongs to which agent and towards which one. The colors of the nodes represent the state of the agents at that time, red as defector and green as cooperator.}
\label{Star}
\end{figure*}

The manner in which the network develops over time is schematically depicted in  Figure~\ref{Star}. This figure shows the recorded connection propensities between 20 $SA$ agents at three time periods. Intensity of the lines between pairs show the magnitude of the propensity of one to connect to the other and the directions show the intensity belongs to which agent and towards which one.

The colors of the nodes represent the state of the agents at that time, red as defector and green as cooperator. The figure shows the connections among the agents at $t=10^{1}$ (left panel), passing through an intermediate state (middle panel, $t=10^{3}$) and after reaching dynamic equilibrium (right panel, $t=10^{5}$). The preferential connections forming the dynamic network emerge here over time and are based on the perception of the benefit that an agent receives from other agents, with whom it interacts. Some connections become stronger whereas others become weaker according to the $SAC$ mechanism. In \cite{mahmoodi19} we showed that $SAC$ creates a complex temporal network with an inverse power law ($ IPL$) probability density function ($ PDF$) $1/p^{\beta } $ of the propensity $p$ of making connections between agents with $ IPL$ index $\beta =1.3$. The $ IPL$ $ PDF$ is very different from the Poisson $ PDF$, the latter having an exponentially diminishing probability of changing the propensity compared with the much greater $ IPL$ value. Consequently, the propensity of forming a link between partners is much greater for the $ IPL$ and the $SAC$ forms a much more complex network than does a Poisson $ PDF$.

In the next section we study the adaptability of social systems ruled by different steps of the $SA$. Disrupting these systems is done by changing some agents to zealots and tracking the changes among the remaining agents in response to the zealots.

\subsection{Complex adaptation of a selfish organization}
To investigate the dynamics of the $SA$ agents we disrupt the stable behavior pattern emerging from the social group by fixing a fraction $f = 0.5$ of the $N=20$ agents to be zealots and calculating the response of the remaining agents. A zealot is an agent whose properties are: zero tendency to be a Cooperator; zero trust to other agents; and a uniform tendency to play the $ PDG$ with other agents. This divides the system into two subsystems: a fraction $f = 0.5$ of the agents that continue to evolve  based on $SAL$, $SALT$, $SALC$ or $SALTC$ (subsystem $S$) and $(1-f)N$ agents as zealots (subsystem $\overline{S}$). We investigate how the remaining $fN$ agents of system $S$ adapt to the new environment by various modes of learning. The degree to which these agents can sustain their mutual cooperation in the presence of the zealots is a measure of their resilience, or its compliment is a measure of their fragility, a fundamental property of complex networks subject to perturbation.

To track the behavioral changes of the agents in $\overline{S}$ from those in $S$ we study the chance of an event happening within a given $SA$ cycle. This could be the chance of finding the pairings Cooperation-Cooperation, Trust-Trust, etc. In previous sections we used the ratio of events, which was useful as there was no perturbation in the systems to detect. To evaluate the chance of the event occurring we used ensemble averages over $10^{3}$ realizations of each simulation.

\begin{figure*}[th]
\centering
\includegraphics[width=14cm]{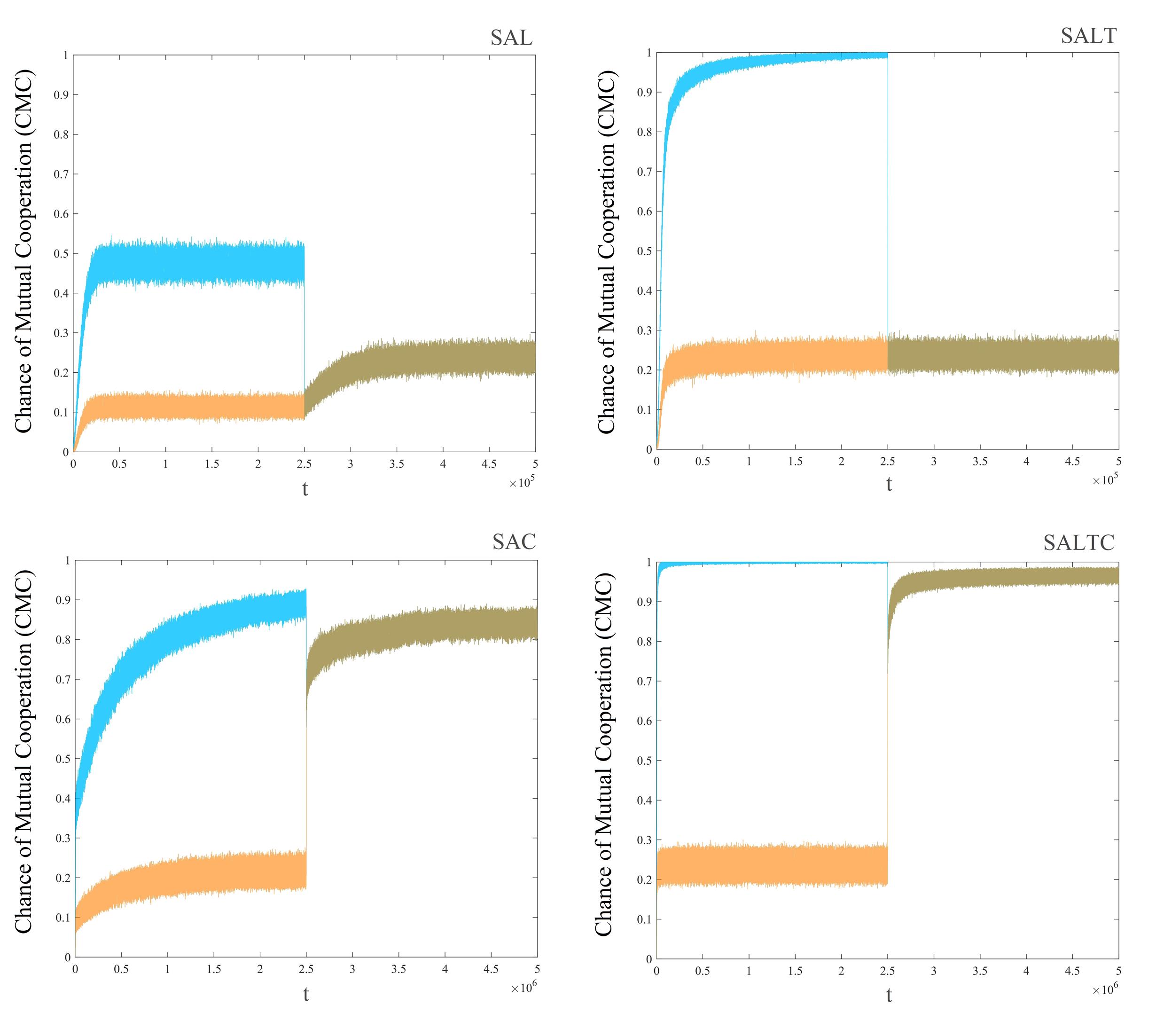}
\caption{The blue curves show the time evolution of the Chance of Mutual Cooperation ($CMC$) (chance of CC to happen among 20 agents at time t) and the orange curves shows the $CMC$ between 10 agents who didn't forced to be zealot. In top-left panel agents used $SAL$ and in top-right panel used $SALT$ to update their decisions while since $t_{z} = 2.5\times10^5$ 10 of the agents forced to be zealots. In bottom-left panel the agents used 
$SALC$ and in bottom-right panel used $SALTC$ to update their decisions while since $t_{z} = 2.5\times10^6$ 10 of the agents forced to be zealots. $N = 20$, $T_{c} = 0.9$. }
\label{CCZealots}
\end{figure*}

The blue and orange curves in the panels of Figure~\ref{CCZealots} show the $CMC$ within a group of $N = 20$ agents and $CMC$ within the subsystem of the 10 agents ($CMC_{S}$) who were not exchanged with zealots after time $t_z$. The top-left panel, shows the $CMC$ and $CMC_{S}$ between the agents which used $SAL$ to update their decisions, before and after 10 of them  being replaced by zealots at time $t_z=2.5\times 10^{5}$. There is a drop in the $CMC$ because of the inevitable pairings between the SAL-agents and the zealots. The $CMC_{S}$ between the 10 agents who were not switched with zealots increased after their interactions with the zealots, despite the overall decrease in the $CMC$ in the system. In other words, the agents of the subgroup $S$ improved their mutual cooperation from about 0.1 to about 0.25. This is because when the agent of subsystem $S$ is randomly paired with a zealot of the subsystem $\overline{S}$, with high probability, it ended up as a sucker and received the minimum payoff of zero. The agent used this payoff as a measure on which to base its  next decision. When an agent of subsystem $S$ paired with another agent of the same subsystem, with high chance (because of their past experience of playing together) played $C$, but because of the low payoff it received previously playing with a zealot, still increases its tendency to play $C$ with this agent.

The top-right panel in Figure~\ref{CCZealots} depicts the $CMC$ and $CMC_{S}$ between agents used $SALT$ to make decisions, before and after the switching time $ t_{z} = 2.5\times10^5$. Although $SALT$ highly increased the $CMC$ level, it failed to sustain that level due to the influence of the zealots. After the switching time $t_{z}$ only the agents of the subsystem $S$ contributed to the mutual cooperation. 

The bottom-left panel in the figure depicts the advantage of adding the $SAC$ to the $SAL$ mechanism in sustaining the $CMC$ of the system after the switching time $t_{z} = 2.5\times10^6$.  This figure shows that after the time $t_{z}$ the $CMC_{S}$ of the subsystem $S$ increases, as the only contribution for mutual cooperation, and saturates at about 0.85. 
The bottom-right panel of the figure shows the $CMC$ and $CMC_{S}$ of the agents use $SALTC$ (entire $SA$ algorithm) to update their decisions. The panel shows a very high resilience of the system even after this massive number of SA-agents switched to zealots at time $t_{z} = 2.5\times10^6$. Similar to the system governed by $SALC$, there is a drop in the $CMC$ but the agents could sustain the level of $CMC_{S}$ to about 0.95. This is because the SA-agents of this system learned to disconnect themselves from the zealots using the $SAC$ mechanism, which highly increased the robustness of the emerging mutual cooperation. Notice that after the switching time $t_{z}$ all the mutual cooperation occurs within subsystem $S$.  

\begin{figure}[bt]
\centering
\includegraphics[width=8cm]{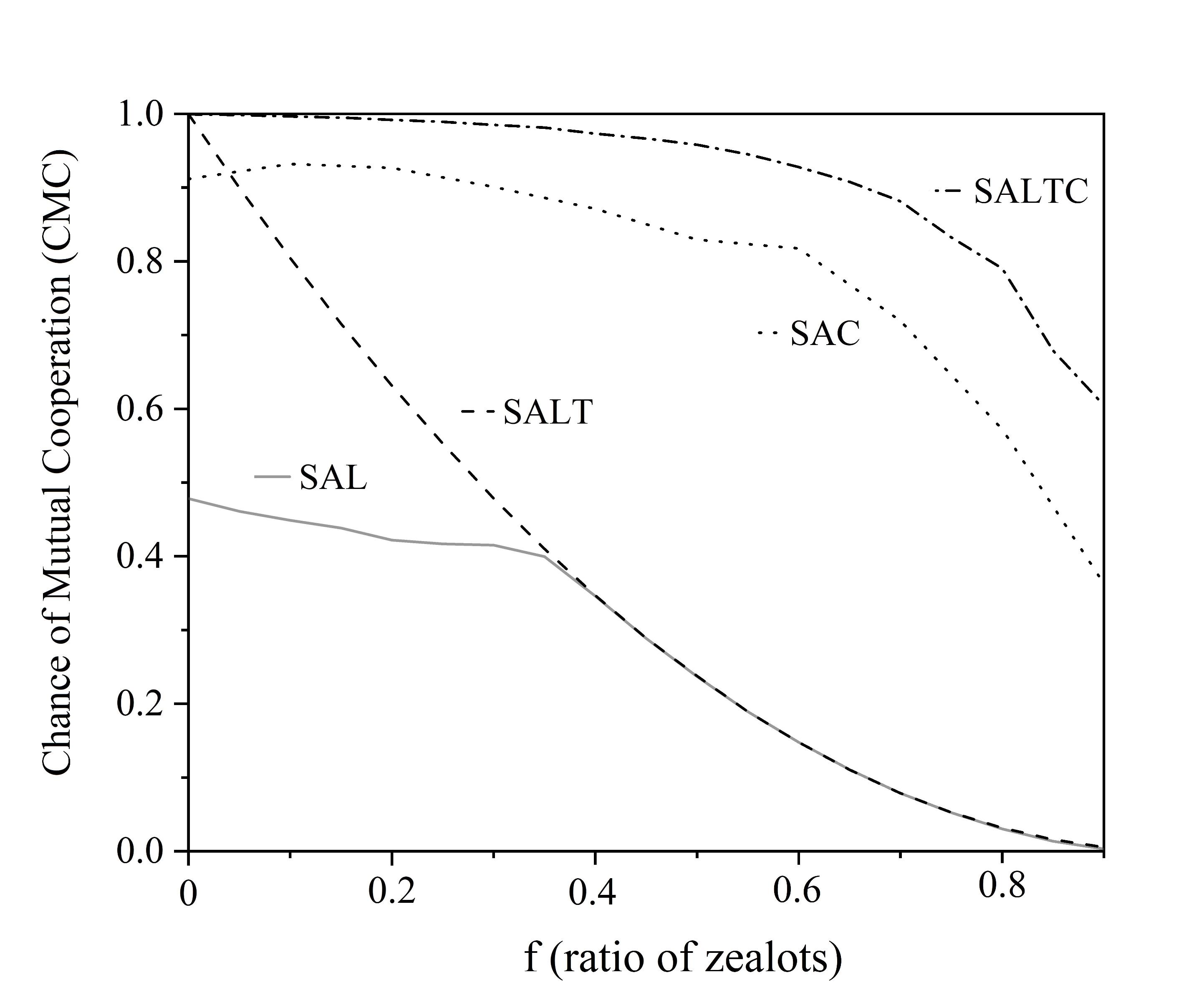}
\caption{Effect of ratio of the agents turned to the zealots at $t_{z} = 2.5\times10^6$ (x axis) on the saturated value of the Chance of Mutual Cooperation ($CMC$) of the remained agents evolved using different steps of the $SA$ algorithm; $SAL$ (solid curve), $SALT$ (dashed curve), $SALC$ (dot curve) and $SALTC$ (dot-dash curve). $N = 20$, $T_{c} = 0.9$.}
\label{Zealots}
\end{figure}

Figure~\ref{Zealots} summarizes the influence a given fraction of zealots within a group manifest under the different learning mechanisms. The figure shows the saturated value of $CMC_{S}$ of the $SA$ agents, using different steps of the $SA$ algorithm for decision making, after a fraction $f$ of the agents turned to zealots at time $t_{z}$. The more adaptable the system, the higher the saturation value of $CMC_{S}$ that can be achieved by the remaining (1-$f$) agents. This provides us with a measure of the resilience of the system. The solid curve is the saturated $CMC_{S}$ for the agents using $SAL$ as a function of the fraction (f) of the agents who switched to zealots at time $t_{z}$ . For $f$ < 0.4 (8 zealots) the system retains a $CMC_{S}$ slightly above 0.4. Beyond this fraction of zealots there is an exponential decay of $CMC_{S}$. In this situation the $SA$ agents learn to modify their decisions ($C$ or $D$) depending on whether or not their pair is a zealot or another $SA$ agent using $SAL$ to make its decisions. 

The dashed curve in the figure  is the saturated value of the $CMC$ of the $SA$ agents, using $SALT$,  after the switching time $t_{z}$. Introducing the $SAT$ learning improves the resilience of the system for $f < 0.4$ above that of $SAL$ alone. However, beyond $f = 0.4$ the calculation converges with the earlier one indicating that the additional learning mechanism of trust ceases to be of value beyond a specific level of zealotry. 

The two top curves of Figure~\ref{Zealots}  are saturated $CMC$ for the $SA$ agents using $SALC$ (doted curve) and $SALTC$ (dot-dash curve) to make decisions, after the switching time $t_{z}$. These two curves show substantial improvement in the system's resilience to an increase in the fraction of zealots. Or said differently, the robustness of the system to perturbations is insensitive to the number of zealots, that is, until the zealots constitute the majority of group membership. The $SAC$ learning in the decision making of the SA-agents have the ability to avoid pairing with the zealots. Once the fraction of zealots exceeds 0.6 the resilience drops rapidly. The saturated $CMC$ for the $SALTC$ is highest where the three learning levels (decision, trust, connection) are active, giving the SA-agents maximum adaptability, that is, flexibility to change in order to maximize their self interest. This realization of self-interest improves the payoff of the whole system as well.

Notice that as the number of the zealots increases, it takes longer for agents to agree to play, but when they do play, they do so with a high propensity that they will cooperate with each other. In other words, increasing the number of zealots slows down the response of the system.

In Appendix B we show complementary analyses of the evolution of the propensity of an agent to interact with other agents after zealots are introduced, as well as the snapshots of the propensities of connections among the 20 agents.

\section{Discussion and Implications of Results}

The Selfish Algorithm provides a demonstration that the benefit to an individual within a social group need not be achieved at the cost of diminishing the overall benefit to the group. Each individual within the group may act out of self-interest, but the collective effect can be quite different than what is obtained by the naive linear logical extrapolation typically made in the tragedy of the commons arguments \cite{lloyd33}. In fact, the $SA$ demonstrates how robust cooperation can emerge out of the selfishness of the individual members of a system that improve the performance of each agent, as well as, that of the overall group, resolving the altruism paradox.

A collective group intelligence emerges from the $SA$ calculations, one based on the self-interest of the individual members of the group.  This collective intelligence grows spontaneously in the absence of explicit awareness on the part of individuals of the outcomes of the behavior of other members of the group. Perhaps more importantly, the collective intelligence develops without assuming a pre-existing network structure to support the collective behavior, and without assuming knowledge of information about the actions or outcomes from other members of the society.

As demonstrated in this research, collective intelligence entailed by the $SA$ unfolds as a consequence of three learning mechanisms:

$SAL$ supports reinforced learning based on the selfishness of agents who play the $PDG$. An agent chooses $C$ or $D$ in its play with other agents and agents influence one another through their payoffs, resulting in the emergence of mutual cooperation, that reveals a collective intelligence.

$SAT$ tunes the propensity of an agent to imitate the strategy of their partner. The strength of the tuning is proportional to the trust being beneficial or detrimental to the agent itself. The role of $SAT$ is to assist agents in achieving coordination, which results in a reduction in the formation of $CD$ pairs. This reduction makes it easier for the system to learn to select $CC$ over $CD$ over time.

$SAC$ tunes the propensity of an agent to connect to other agents based on how they improved its self-interest.

We have studied the advantage of each of the $SA$ learning mechanisms in creating mutual cooperation and thereby the facilitation of the growth of group intelligence. We also studied the adaptation and resilience of the emergent intelligence by replacing some of the agents with zealots and examining the system's response. The control of the dynamics in $SA$ is internal and according to the self-interest of the agents, spontaneously directs the system to robust self-organization. The most robust systems use the $SAC$ mechanism to make decisions, which increase the complexity of the system by forming a dynamic complex network with a high propensity for partnering, thereby providing the agents with the ability to learn to shun the zealots. These complex temporal pairings that emerge from $SAC$ can be seen as a source of the network reciprocity introduce by Nowak and May \cite{nowak1992evolutionary}.

One advantage of the $SA$ is its simplicity which enables us to introduce additional behavioral mechanisms into the dynamics such as deception or memory. This makes it possible to create algorithms for different self-organizing systems such as those used in voting. $SA$ leadership emerges anywhere and everywhere as the need for it arises and it leads to socio-technical systems which achieve optimum leadership activities in organizations to make teams more effective at achieving their goals. The model also has promise in the information and communication technology (ICT) field of sensor and sensor array design. Future research will explore the application in detecting deception in human - human or human - machine interactions and in anticipating threats or providing leaders with near real-time advice about the emerging dynamics of a given specific situation.

The predictions in this work are entirely simulation-based, however a number of them are being experimentally tested. Experimental work (paper in preparation) confirms the emergence of mutual cooperation between human pairs playing PD when they are given the option to trust their pair's decision, in no explicit awareness of the payoffs to other agents. Additional experiments to test our hypothesis that the collective cooperation emerged by $SA$ can survive to perturbations are in progress.

\section*{Acknowledgments}

This research was supported by the Army Research Office, Network Science Program, Award Number: W911NF1710431.

\bigskip

\section*{Appendix A: Details for the $SA$ algorithm}

\begin{figure*}[bt]
\centering
\includegraphics[width=14cm]{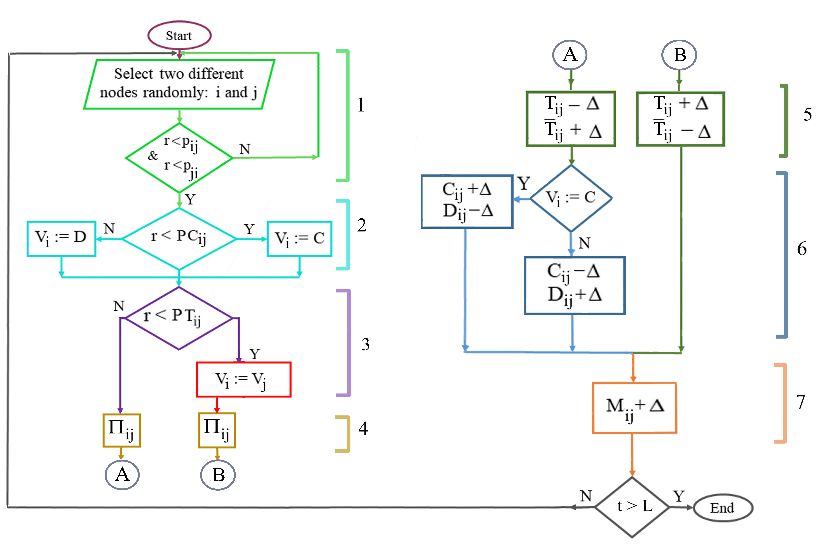}
\caption{Flowchart of the $SA$. "Y" and "N" letters represent "Yes" and "No", respectively.} 
\label{Flowchart}
\end{figure*}

The following steps are executed during a single cycle initiated at time $t$ of the $SA$ algorithm depicted in Figure~\ref{Flowchart} \qquad \qquad \qquad

\paragraph{1. Selfish algorithm - connection ($SAC$)}

Agents $i$ and $j$ are picked randomly such that at time $t$ the $SAC$ agent $ i $ has the propensity: 
\begin{equation}
P_{ij}(t)=\frac{M_{ij}(t)}{\sum_{k}M_{ik}(t)}  \label{$SAC$}
\end{equation}
to play with agent $j$, where the propensity falls in the interval $
0<P_{ij}(t)<1$. The quantity $M_{ij}(t)$ is the cumulative tendency for agent $i$ to select agent $j$ to play at time $t$. This cumulative tendency changes at step 7, according to the last two payoffs received by agent $i$.

At the same time, the $SAC$ agent $j$ has a propensity given by Eq. (\ref{$SAC$}), with the indices interchanged, to play with agent $i$, where the propensity falls in the interval $0<P_{ji}(t)<1$. Two agents $i$ and $j$ partner if two numbers $r_{1}$ and $r_{2}$ are randomly chosen from the interval (0,1), and satisfy inequalities $r_{1}<P_{ij}(t)$ and $ r_{2}<P_{ji}(t)$. If both inequalities are not satisfied another two agents are randomly selected at each time $t$ until the inequalities are satisfied and the two agents "agree" to play a $ PDG$. In the flowchart wherever letter $r$ is called it returns a single uniformly distributed random number in the interval (0,1).

\paragraph{2. Selfish algorithm - learning ($SAL$)}

Agent $i$, playing with agent $j$, initially selects an action, $C$ or $D$, using $SAL$. The agent $i$ has the propensity: 

\begin{equation}
PC_{ij}(t)=\frac{C_{ij}(t)}{C_{ij}(t)+D_{ij}(t)}  \label{CSAL}
\end{equation}
to pick $C$ and the propensity:
\begin{equation}
PD_{ij}(t)=\frac{D_{ij}(t)}{C_{ij}(t)+D_{ij}(t)}  \label{DSAL}
\end{equation}
to pick $D$, as it's next potential decision. Note that the sum of these two terms is one. The quantities $C_{ij}(t)$ and $D_{ij}(t)$ are cumulative tendencies for agent $i$ playing with agent $j$, at time $t$, for the choice  $C$ or $D$, respectively. These cumulative tendencies change at step 6 based on the last two payoffs of agent $i$.

To decide on an action a random number $r$ is selected from the interval (0,1). If $r<PC_{ij}(t)$ then the next decision of $SAL$ agent $i$ will be $C,$ otherwise will be $D$. The same reasoning applies for agent $j$.

\paragraph{3. Selfish algorithm - trust ($SAT$)}

Instead of executing the decision determined by $SAL$ in step 2, agent $i$ has a propensity to trust the decision made by agent $j$, which also used $SAL$ in step 2. The propensity for agent $i$ to trust the decision of agent $j$ is:
\begin{equation}
PT_{ij}(t)=\frac{T_{ij}(t)}{T_{ij}(t)+\overline{T}_{ij}(t)}.  \label{$SAT$}
\end{equation}
Here again, if a random number $r$, chosen from the interval (0,1), is less
than $PT_{ij}(t)$ then trust is established. The quantities $T_{ij}(t)$ and $\overline{T}_{ij}(t)$ are cumulative tendencies for agent $i$ to execute the choice of agent $j$, at time $t$, or to not rely on trust and to execute its choice based on $SAL$, respectively. These cumulative tendencies update in step 5 based on the last two payoffs of agent $i$.

\paragraph{4. Evaluating Own Payoffs}

After agent $i$ and $j$ executed their action, $C$ or $D$, their payoff is evaluated using the payoffs matrix of the $ PDG$, $\Pi _{ij}(t)$ and $\Pi_{ji}(t)$, respectively.

\paragraph{5. Update of cumulative tendency of $SAT$}

If agent $i$ used $SAT$ playing with agent $j$ then the accumulative tendencies $T_{ij}$ and $\overline{T}_{ij}$ change to $T_{ij}+\Delta _{i,jk}$ and $\overline{T}_{ij}-\Delta _{i,jk}$ for the next time agent $i$ and $j$ partner. The same happens for agent $j$. Similarly, if agent $i$ did not use $SAT$, but relied on its own $SAL$,  then the accumulative tendencies $\overline{T}_{ij} $ and $T_{ij}$ change to $\overline{T}_{ij}+\Delta _{i,jk}$ and $T_{ij}-\Delta _{i,jk}$ for the next time agent $i$ and $j$ partner. The same happens for agent $j$.

\paragraph{6. Update of cumulative tendency of $SAL$}

Step 6 is only active if the agent did not use $SAT$ at step 3. If agent $i$ played $C$ with agent $j$, then the accumulative tendencies $C_{ij}$ and $D_{ij}$ change to $ C_{ij}+\Delta _{i,jk}$ and $D_{ij}+\Delta _{i,jk}$ for the next time agent $i$ and $j$ partner. If agent $i$ played $D$ with agent $j$, then the accumulative tendencies $D_{ij}$ and $C_{ij}$ change to $ D_{ij}+\Delta _{i,jk}$ and $C_{ij}-\Delta _{i,jk}$ for the next time agent $i $ and $j$ partner. The same happens for agent $j$.

\paragraph{7. Update of cumulative tendency of $SAC$}

In this step the cumulative tendency to play with a specific agent changes. If agent $i$ played with agent $j$ then the cumulative tendency of pairing with agent $j$, $M_{ij}$ changes to $M_{ij}+\Delta _{i,jk}$. The same happens for agent $j$.

As boundary condition for steps 5, 6 and 7, if the updated  cumulative tendency goes beyond its defined maximum  (or below its defined minimum) then it has to set back to the maximum value (minimum value).

\section*{Appendix B: Effect of zealots on the dynamics of the SA}

\begin{figure*}[bt]
\centering
\includegraphics[width=15cm]{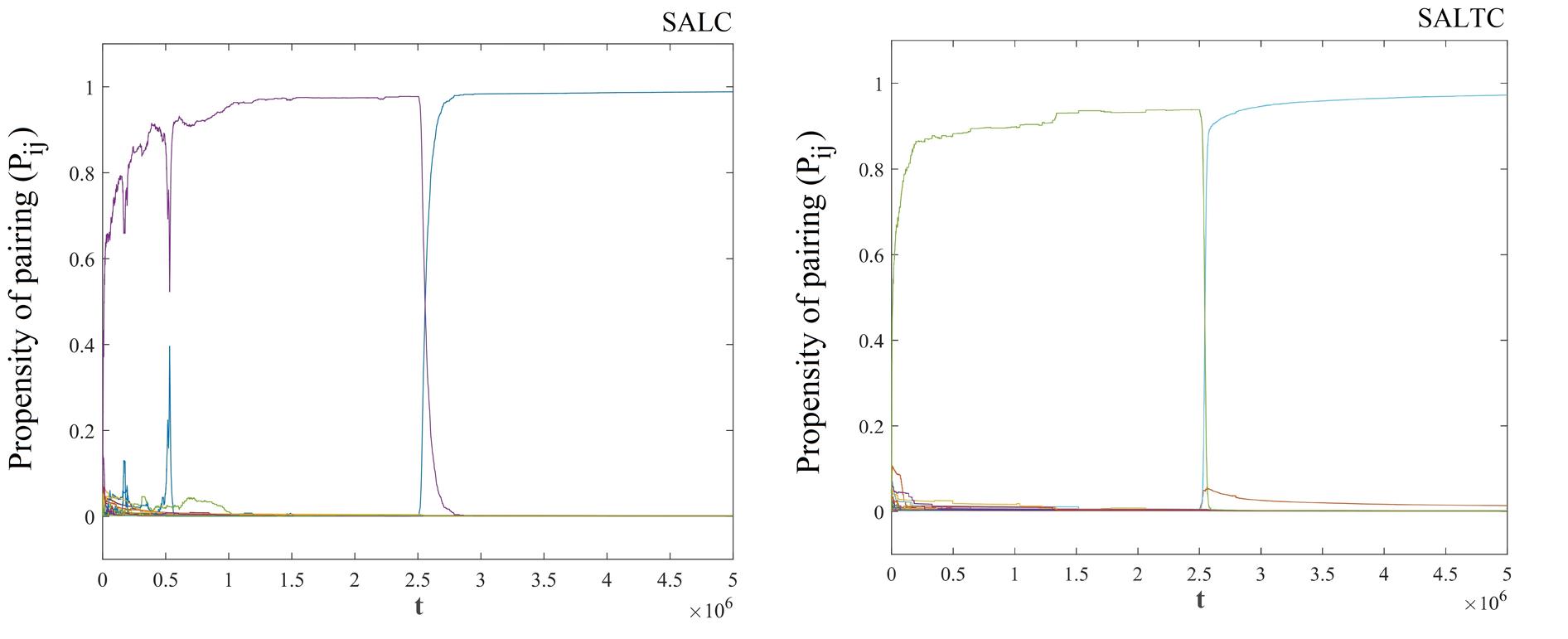}
\caption{Time evolution of the propensity  of agent 1 to pair with the other 19 agents $P_{1j}$, where agents used $SALC$ (left panel) and $SALTC$ (right panel) before $t_{z} = 2.5\times10^6$ and after $t_{z}$ 10 of the agents (out of 19) turned to zealots. $N = 20$, $T_{c} = 0.9$.}
\label{timeZealotLTC}
\end{figure*}

Figure~\ref{timeZealotLTC} shows the time evolution of the propensity of $SAC$ of one agent 1 to other 19 agents, in system of 20 agents which used $SALT$ (left panel) or $SALTC$ (right panel) to make decision until $t_{z}$ where 10 (out of the 19) agents turned to zealots. This figure shows that agent 1, before $t = t_{z}$, mostly played with the agent corresponding to the purple or green curve of left or right panel respectively. But after these agents switched to zealots then agent 1 decreased its propensity to pair up and play with them and switched to pair with the agent corresponding to blue curves (not a zealot).

\begin{figure*}[bt]
\centering
\includegraphics[width=18cm]{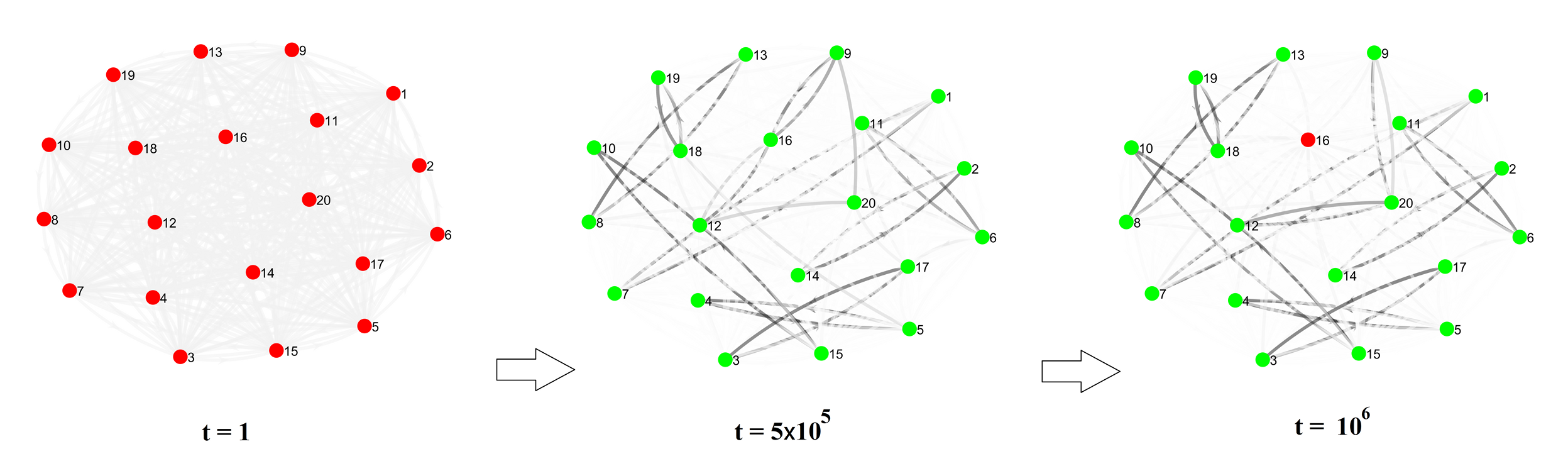}
\caption{Complex adaptation in the dynamic network emerged by SA. From left to right the panels show the snapshots of the propensities of connections between 20 agents at $t= 1$, $t= 5\times10^{5}$ and $t=10^{6}$, respectively. Agents played the $ PDG$ and updated their decisions based on $SALTC$ except agent 16 (red node on the left panel) which turned to a zealot and kept  defecting after $t_{z} = 5\times10^{5}$. Intensity of the lines between pairs represent the magnitude of the propensity of one to another and the directions show the intensity belongs to which agent and towards which one. The colors of the nodes represent the state of the agents at that time, red as defector and green as cooperator. $M = 20$, $T_{c} = 0.9$.}
\label{AnimZealot7}
\end{figure*}

Figure~\ref{AnimZealot7} shows the snapshots of the propensities of connections between 20 agents, using $SALTC$ to make decisions, before and after one of them turned to a zealot (agent 10). This figure shows that the network is intelligent and is able to isolate the zealot, in order to sustain the highest level of mutual cooperation possible.

\section*{Appendix C: Effect of trust (fixed chance of trust) on the evolution of decisions}

\begin{figure*}[bt]
\centering
\includegraphics[width=15cm]{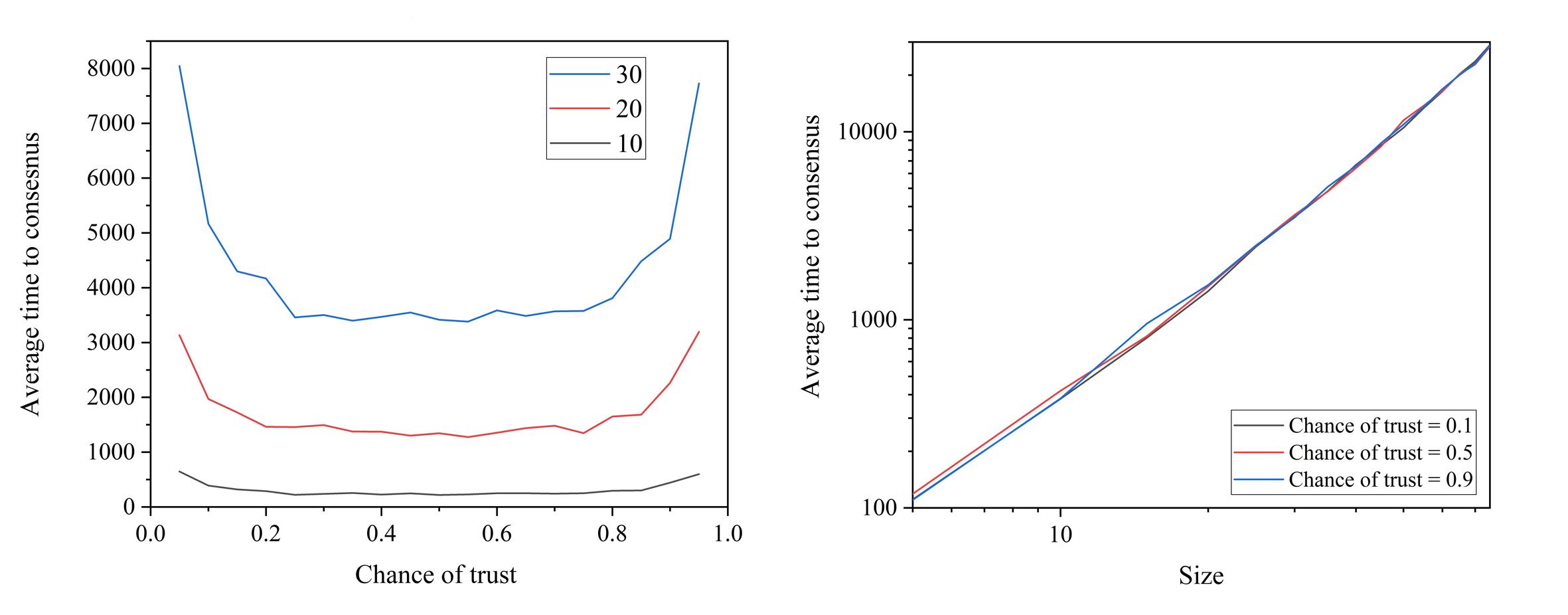}
\caption{Left panel: The black, red and blue curves show the average time to reach consensus (the time that all the agents reach in +1 or all in -1 state) vs. the magnitude of the trust between them for systems with 10, 20, and 30 agents, respectively. At each time an agent can keep its previous decision or can imitate the decision of its partner by the fixed propensity of trust. Right panel: Dependence of the average time to consensus to the number of the agents with 0.1 (black), 0.5 (red) and 0.9 chance of trusting the decision of their pair. The power law coefficients for $N^{\alpha}$ are 2.37, 2.28 and 2.29, respectively. The curves are ensemble averages over 100 realizations.}  

\label{agreement}
\end{figure*}

The aim of this section is to highlight the difference between trust used in the literature with our dynamic trust. We introduced "trust" and "not trust" as a "decision" (while the word trust means "not making a decision"). We connected trust to a reinforcement learning mechanism to make it adaptable.  To show the difference, here we determine the effect of fixed chance of trust between the agents on the dynamics of decisions where there is no reinforcement learning active, which is to say that trust is the only learning mechanism through $SAT$. We investigated the dynamics of the decisions made by social groups consisting of 10, 20, or 30 agents. These agents randomly partner at time $t$ and can either retain their decision, or exchange it with a fixed chance for their partner's decision (trusting). In left panel of Figure~\ref{agreement} the average time for the agents to reach consensus is plotted versus the chance of trust that agents have to take the decision of their partner as their next decision in each of the three groups. This figure shows that there is a broad flat minimum (optimum) value of trust that leads the agents to achieving consensus faster and that this time increases with the size of the network.

It is interesting to note that the evolution of mutual cooperation when one of the agents becomes a zealot, which is to say, that agent remains the same all the time, the network rapidly relaxes to the value of the zealot. The conclusion is that this network is highly sensitive to this small perturbation depending on the chance of trust. In other words, this system, that learns only through trust is not significantly resilient (is not robust).

The panel on the right of the Figure~\ref{agreement} indicates a dependence of the average time for a group to reach consensus from a random initial state  is a monotonously increasing function of group size N. The average time to reach consensus satisfies an allometry equation \cite{west2017nature}:  $Y=aX^{b}$, where Y is the functionality of the system, X is a measure of system size and the empirical parameters a and b are determined by data. Allometry relations (ARs) have only recently been applied to social phenomena \cite{bettencourt2013origins} but on those applications the scaling has always been superlinear (b>1). Bettencourt  \cite{bettencourt2013origins} point out that the superlinear scale reflects unique social characteristics with no equivalent in biology in which the ARs scaling is invariably sublinear (b<1). They point out that the knowledge spillover in urban settings drive the growth of functionality such as wages, income, gross domestic product, bank deposits, as well as rates of invention...all scale superlinearly with city size.

In this paper  the agents were allowed to learn through a variety of modalities and to find their individual optimum value of trust regarding other agents. Subsequently, we showed that this dynamic trust acting, in concert with other learning modalities, leads to robust mutual cooperation.


\begin{references}




\bibitem{nowak1993strategy} Nowak M, Sigmund K, \emph{A strategy of win-stay, lose-shift that outperforms tit-for-tat in the Prisoner's Dilemma game}, Nature, {\bf 364}, 56--58 (1993). 


\bibitem{gonzalezetal2015} Gonzalez C, Ben-Asher, N Martin JM, Dutt V, \emph{A Cognitive Model of Dynamic Cooperation With Varied Interdependency Information}, Cognitive Science, {\bf 39}, 457-495 (2015). 



\bibitem{west17} Mahmoodi K, West BJ, Grigolini P, \emph{Self-organizing complex networks: individual versus global rules}, Frontiers in physiology, {\bf 8}, 478 (2017). 



\bibitem{lloyd33} Lloyd WF, \emph{Two Lectures on the Checks to Population: Delivered Before the University of Oxford, in Michaelmas Term 1832}, JH Parker (1833). 


\bibitem{ariely08} Ariely D, \emph{Predictably irrational: The hidden forces that shape our decisions}, HarperCollins (2008). 




\bibitem{kahneman2011thinking} Kahneman D, \emph{Thinking, fast and slow}, Macmillan (2011). 



\bibitem{darwin71} Darwin C, \emph{The Origin of the Species and the
Descent of Man}, Modern Library, NY (1871). 



\bibitem{wilson07} Wilson DS, Wilson EO, \emph{ethinking the theoretical foundation of sociobiology}, The Quarterly review of biology, {\bf 4}, 327--348 (2007). 





\bibitem{west19b} West BJ, Mahmoodi K, Grigolini P, \emph{Empirical Paradox, Complexity Thinking and Generating New Kinds of Knowledge}, Cambridge Scholars Publishing (2019). 





\bibitem{rand2013human} Rand DG, Nowak, Martin A, \emph{Human cooperatio}, Trends in cognitive sciences, {\bf 17},413--425 (2013). 



\bibitem{RapoportChammah65} Rapoport A, Chammah AM, \emph{Prisoner's Dilemma: A Study in Conflict and Cooperation},University of Michigan Press (1965). 




\bibitem{ Moisanetal2018} Moisan F,  ten Brincke R, Murphy RO, Gonzalez C, \emph{Not all Prisoner's Dilemma games are equal: Incentives, social preferences, and cooperation}, Decision, {\bf 5}, 306--322 (2018). 




\bibitem{nowak1992evolutionary} Nowak MA, May RM, \emph{Evolutionary games and spatial chaos}, Nature, {\bf 359}, 826 (1992). 



\bibitem{MartinGonJuvLeb2014} Martin JM, Gonzalez C, Juvina I, Lebiere C, \emph{A Description-Experience Gap in Social Interactions: Information about Interdependence and Its Effects on Cooperation}, Journal of Behavioral Decision Making, {\bf 27}, 349-362 (2014). 



\bibitem{fischbacher2001people} Fischbacher U, G{\"a}chter S, Fehr E, \emph{Are people conditionally cooperative? Evidence from a public goods experiment},Economics letters, {\bf 71}, 397--404 (2001). 




\bibitem{barabasi1999emergence} Barab{\'a}si AL, Albert R, \emph{Emergence of scaling in random networks}, Science, {\bf 286}, 509--512 (1999). 



\bibitem{tomassini2007social} Tomassini M, Pestelacci Enea, Luthi L, \emph{Social dilemmas and cooperation in complex networks}, International Journal of Modern Physics C, {\bf 18}, 1173--1185 (2007). 



\bibitem{adami2016evolutionary} Adami C, Schossau J, Hintze A, \emph{Evolutionary game theory using agent-based methods}, Physics of life reviews, {\bf 19}, 1--26 (2016). 



\bibitem{couzin2007collective} Couzin I, \emph{Collective minds}, Nature, {\bf 445}, 715 (2007). 



\bibitem{li2018ising} Li C, Liu F, Li P, \emph{Ising model of user behavior decision in network rumor propagation}, Discrete Dynamics in Nature and Society, {\bf 2018},  (2018). 






\bibitem{mahmoodi19} Mahmoodi K, Gonzalez C, \emph{Emergence of collective
cooperation and neworks from selfish-trust and selfish-connections}, Cognitive Science (The 41st Annual Meeting of the Cognitive Science Society), https://mindmodeling.org/cogsci2019/papers/0392/0392.pdf, 2254--2260 (2019). 



\bibitem{gintis2009bounds} Gintis H, \emph{The Bounds of Reason: Game Theory and the Unification of the Behavioral Sciences-Revised Edition}, Princeton University Press (2014). 




\bibitem{Gonzalezetalinprep} Gonzalez C, Mahmoodi K, Singh K, \emph{Emergence of Collective Cooperation in the Absence of Interdependency Information}, (in preparation) 2020. 




\bibitem{west2017nature} West BJ, \emph{Nature’s Patterns and the Fractional Calculus}, Walter de Gruyter GmbH \& Co KG (2017). 



\bibitem{bettencourt2013origins} Bettencourt LMA, \emph{The origins of scaling in cities}, Science, {\bf 340}, 1438--1441 (2013). 







 \end{references}
\end{document}